\documentclass[acmtog,nonacm]{acmart}

\AtBeginDocument{%
  \providecommand\BibTeX{{%
    \normalfont B\kern-0.5em{\scshape i\kern-0.25em b}\kern-0.8em\TeX}}}

\usepackage{xcolor}
\usepackage{soul}
\definecolor{good}{HTML}{66bb6a}
\definecolor{inc}{HTML}{d4e157}
\definecolor{mid}{HTML}{ffca28}
\definecolor{bad}{HTML}{ff7043}

\definecolor{todotext}{RGB}{0,0,0}
\definecolor{todobg}{RGB}{255, 219, 168}

\usepackage{caption}
\usepackage{subcaption}

\usepackage{multirow}

\usepackage{rotating}

\usepackage{colortbl}

\begin{document}

\title{A Multi-Layout Design for Immersive Visualization of Network Data}

% \affiliation{%

\author{David Bauer}
\email{davbauer@ucdavis.edu}
\affiliation{%
  \institution{UC Davis}
  \country{United States}}

\author{Chengbo Zheng}
\affiliation{%
  \institution{HKUST}
  \country{China}}
\email{cb.zheng@connect.ust.hk}

\author{Oh-Hyun Kwon}
\email{kw@ucdavis.edu}
\affiliation{%
  \institution{UC Davis}
  \country{United States}}

\author{Kwan-Liu Ma}
\email{klma@ucdavis.edu}
\affiliation{%
  \institution{UC Davis}
  \country{United States}}

% \affiliation{%

% \affiliation{%

% \affiliation{%

% \affiliation{%

% \affiliation{%

% \affiliation{%


\begin{abstract}
Visualization plays a vital role in making sense of complex network data.
Recent studies have shown the potential of using extended reality (XR) for the immersive exploration of networks. The additional depth cues offered by XR help users perform better in certain tasks when compared to using traditional desktop setups. However, prior works on immersive network visualization rely on mostly static graph layouts to present the data to the user. This poses a problem since there is no optimal layout for all possible tasks. The choice of layout heavily depends on the type of network and the task at hand. We introduce a multi-layout approach that allows users to effectively explore hierarchical network data in immersive space. The resulting system leverages different layout techniques and interactions to efficiently use the available space in VR and provide an optimal view of the data depending on the task and the level of detail required to solve it. To evaluate our approach, we have conducted a user study comparing it against the state of the art for immersive network visualization. Participants performed tasks at varying spatial scopes. The results show that our approach outperforms the baseline in spatially focused scenarios as well as when the whole network needs to be considered.
\end{abstract}

\begin{CCSXML}
<ccs2012>
<concept>
<concept_id>10003120.10003145.10003146.10010892</concept_id>
<concept_desc>Human-centered computing~Graph drawings</concept_desc>
<concept_significance>500</concept_significance>
</concept>
<concept>
<concept_id>10003120.10003145.10003147.10010923</concept_id>
<concept_desc>Human-centered computing~Information visualization</concept_desc>
<concept_significance>300</concept_significance>
</concept>
<concept>
<concept_id>10003120.10003121.10003124.10010866</concept_id>
<concept_desc>Human-centered computing~Virtual reality</concept_desc>
<concept_significance>500</concept_significance>
</concept>
<concept>
<concept_id>10003120.10003121.10003122.10003334</concept_id>
<concept_desc>Human-centered computing~User studies</concept_desc>
<concept_significance>300</concept_significance>
</concept>
</ccs2012>
\end{CCSXML}

\ccsdesc[500]{Human-centered computing~Graph drawings}
\ccsdesc[300]{Human-centered computing~Information visualization}
\ccsdesc[500]{Human-centered computing~Virtual reality}
\ccsdesc[300]{Human-centered computing~User studies}

\keywords{network community structure, graph layout, head-mounted display, immersive environment}

%% 
%% Teaser Figure
\begin{teaserfigure}
    \includegraphics[width=\textwidth]{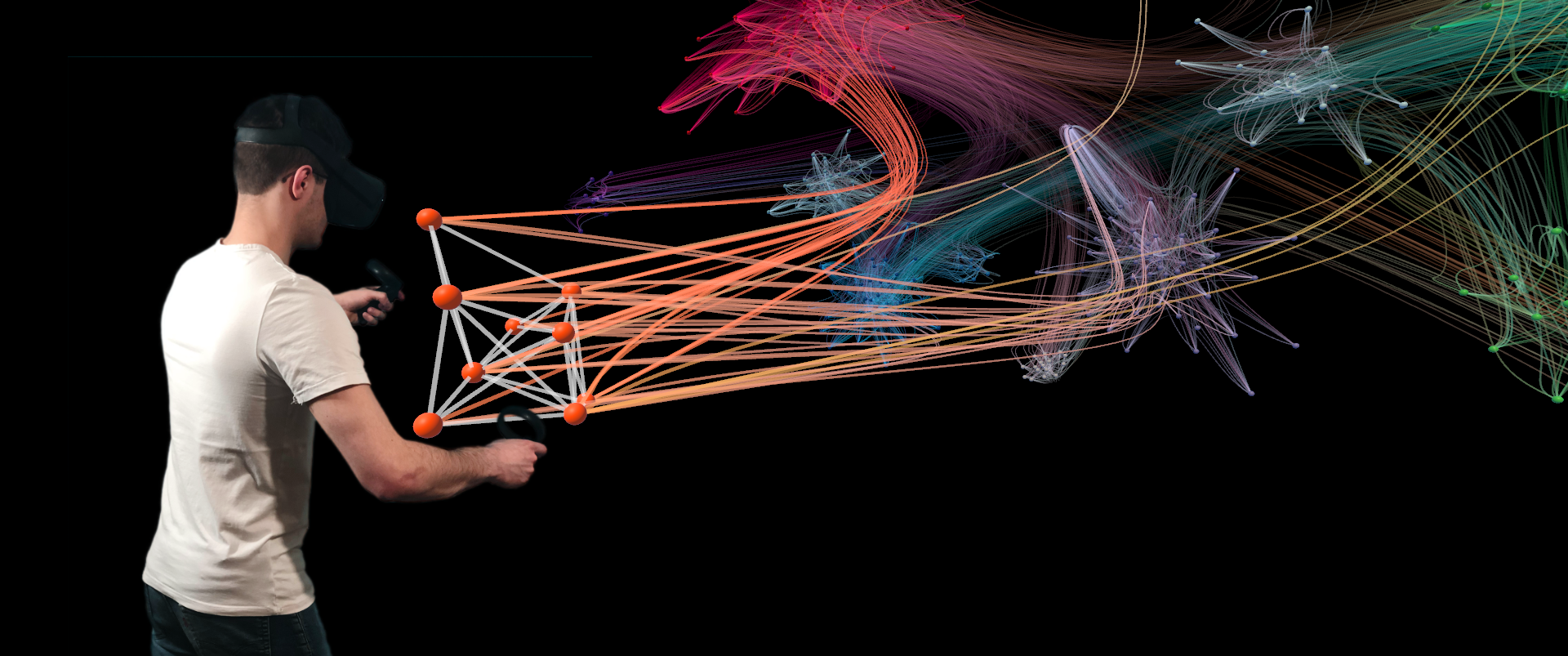}
    \caption{A view of a hierarchical network rendered with our system. A single group is expanded. Its nodes and intra-group connectivity can be inspected in detail while preserving inter-group context.}
    \Description{figure description}
    \label{fig:teaser}
\end{teaserfigure}

\maketitle

%%
%% Section Includes
\section{Introduction}
% {\color{red} Network data, representing relations of interest 
% between entities in a domain under study, is commonly found in 
% a variety areas of study from sociology, computational biology, ecology, to epidemiology and many others.}
Network data represents relations of interest between entities in a domain under study. It is commonly found in a variety of areas such as sociology, computational biology, ecology, to epidemiology and many others.
Building effective visualizations is essential for the analysis and understanding of network data. Many network datasets are quite complex and visualizing them is often the only way to produce insightful analyses. Such networks contain a myriad of nodes and links which themselves form group and sub-group structures within the data. Group structures often arise from similarities in certain node attributes or are formed based on strong connectivity in parts of a network. Depending on the network's semantics, these groups---often called communities---can represent anything from circles of friends in a social network to similar types of bacteria in a biological network. 
% To emphasize this community aspect, it is common practice to encode these structures as a hierarchical network. 
% These hierarchies can be either curated manually or computed automatically. 
% Performing these encodings can be beneficial as there is much to be learned from analyzing communities in a network \cite{Saket2014a}.
% Visualizing a hierarchical network in an intuitive way and providing users with the views on the data necessary to explore this type of data efficiently is crucial for the analysis process. However, providing effective analysis tools for such networks still remains a challenge.

Exploring the community structure of a network using the methods commonly found in most graph drawing disciplines might yield sub-optimal results \cite{Yang2013,Vehlow2017}.
The community structure might not be apparent, and conventional representations and interaction methods might not suffice to provide deeper insight. 
For example, looking at a social network as a conventional node-link diagram, where the nodes are simply connected by lines, might not reveal the intricate interactions between social sub-communities that are inherent to the data.

There have been efforts to cater to analysts' needs by developing layout techniques and visual de-cluttering methods that improve the legibility and highlight the community structure of a network~\cite{Vehlow2017}. 
However, most of the methods are designed for 2D displays. For instance, a common task for group-based network visualization is to find the number of neighboring communities or to identify the group with the highest group-degree (connected to the most other groups)~\cite{Saket2014}. 
Such tasks require users to navigate among multiple communities. In a 2D setting, users need to pan and zoom frequently in large networks due to visual clutter.

Recent years have seen an advance in three-dimensional rendering and layout of networks~\cite{Barahimi2014}. Particularly, stereoscopic displays and head-mounted (HMD) virtual reality (VR) devices and similar immersive technologies were successfully used to facilitate network analysis~\cite{Ware1996,Ware2005,Ware2008}. Although they rarely go beyond simple node-link representations, these works pioneered the field, underscoring the potential of such applications. More recent work has started exploring in-depth the possibilities of network rendering in VR~\cite{Kwon2016}. With the devices that offered these new prospects came a new means of interaction. Motion controls promoted the development of novel interaction designs for virtual environments~\cite{Huang2017}. Even in immersive space, there is no single optimal graph layout or interaction method for all tasks. Despite this fact, prior work has only been focused on rendering networks using single, static layouts with minimal interaction, relying solely on the perceptual advantages of immersive VR to justify their designs.

In our work, we introduce a multi-layout approach to visualizing network data in VR. We created an interactive suite of graph layouts that helps solve a wide variety of tasks and offers the cognitive advantages of immersive VR. Using the space available in VR we are able to present, combine and juxtapose multiple graph layouts---something that is not easily possible in screen-based applications. This kind of approach caters to cognitive requirements that are associated with analysing a network at different scopes. Layouts range from overview summaries of the whole network to detailed projections of single communities - all while visually preserving their context in the dataset. The choice of layouts is informed by the idea of an extended overview-plus-detail workflow. Each layout is designed to be explored from an egocentric perspective. We use several previously established presentation techniques~\cite{Kwon2016,Itoh2009,Didimo2014,Holten2006,Munzner1997,forceatlas2} to create effective and clutter-free visualizations and our interactions let users engage with and navigate between the different layouts.

We conducted a user study comparing our design against the state of the art for immersive network visualization. The results of our evaluation show that although our multi-layout approach might not be suited for all types of community-based tasks, it outperforms the state of the art approach in the majority of cases. Users achieve significantly higher accuracy in the same time in spatially focused tasks as well as in scenarios that require attention to the whole network. Our work is another pivotal step towards establishing virtual reality as an effective means for information visualization.
\section{Related Work}
Our work is related to the use of virtual reality for data visualization, networks with community structures, interaction design and community-based tasks. Here, we discuss existing studies related to these topics.

\subsection{Virtual Reality}
Ever since 1965 when Sutherland et al. have introduced the Ultimate Display~\cite{sutherland1965ultimate}---one of the first virtual reality (VR) devices---the world has seen a multitude of devices and applications of VR. However, its uses have long been confined to highly specialized research and industry applications due to high cost of devices and development~\cite{VanDam2000}. Today, VR has matured and is marketed heavily by the entertainment industry~\cite{VIVETM, PlayStationVR, SamsungGearVR, OculusVR} which has led to the development of cost-effective and widely available devices. 

With the recent advances, the scientific community has also embraced VR as a means for data visualization, education, and training~\cite{Brooks1999}. Early adopters such as NASA have shown the value of using VR in wind tunnel simulations~\cite{Bryson1993}. Later on, it has been used for volume rendering~\cite{Hanel2015, Scholl2018, Scholl2019}, trajectory analysis~\cite{Hurter2019}, diagnostic colonoscopy~\cite{Mirhosseini2015, Mirhosseini2019} and even archaeology~\cite{Bennett2015}, to name a few. We refer the interested reader to Mariott et al.'s~\cite{Marriott2018} detailed analysis of the potential uses of immersive visualizations for more examples. This new domain of research is far from exhausted. As more and more researchers engage with VR, it becomes clear what type of tasks and representations work in VR and which ones do not~\cite{McIntire2015}.
While there exist several recommendations for how VR may be applied to data visualization~\cite{Bryson1993,bowman20043d, VanDam2000}, there are no unified guidelines to do so.

\begin{table*}[!bt]
    \centering
    \caption{Overview of existing work in the domain of immersive graph visualizations. Works are shown in terms of used graph layouts and supported tasks. Task names are based on Lee et al.'s taxonomy of graph tasks~\cite{Lee2006} and Saket et al.'s work on group-level tasks~\cite{Saket2014} and tasks found in cited work have been catorized to best match items in this table. Red cells indicate task-layout combinations that have not been discussed in prior work, while yellow highlights items that have been considered but are not supported by our design. Hues of green indicate items that are supported by our design.}
    \begin{tabular}{@{}llcccc@{}}
        \multicolumn{5}{c}{\textbf{Layouts}}\\
        \cmidrule{2-5}
         & & \textit{2D Force-Directed Layout} & \textit{3D Force-Directed Layout} & \textit{3D Spherical Layout} \\
         \cmidrule{2-5}
         \cmidrule{2-5}
        \multirow{10}{*}{\rotatebox{90}{\textbf{Tasks}}} & \textit{Low-Level} & \cellcolor{inc} \textbf{Ours}, \cite{Alper2011}, \cite{Halpin2008} & \cellcolor{inc} \textbf{Ours}, \cite{Drogemueller2019}, \cite{Huang2017} & \cellcolor{good} \textbf{Ours} \\
        \cmidrule{2-5}
        & \textit{Adjacency} & \cellcolor{inc} \textbf{Ours}, \cite{Alper2011} & \cellcolor{good} \textbf{Ours} & \cellcolor{inc} \textbf{Ours}, \cite{Kwon2016} \\
        \cmidrule{2-5}
        & \textit{Common Connection} & \cellcolor{inc} \textbf{Ours}, \cite{Alper2011}, \cite{Halpin2008} & \cellcolor{inc} \textbf{Ours}, \cite{Huang2017} & \cellcolor{inc} \textbf{Ours}, \cite{Kwon2016} \\
        \cmidrule{2-5}
        & \textit{Connectivity} & \cellcolor{good} \textbf{Ours} & \cellcolor{inc} \textbf{Ours}, \cite{Drogemueller2019}, \cite{Kotlarek2020}, \cite{Bueschel2019}, \cite{Huang2017} & \cellcolor{inc} \textbf{Ours}, \cite{Kwon2016} \\
        \cmidrule{2-5}
        & \textit{Attributes on Nodes} & \cellcolor{mid} \cite{Halpin2008} & \cellcolor{mid} \cite{Drogemueller2019}, \cite{Huang2017} & \cellcolor{bad} --- \\
        \cmidrule{2-5}
        & \textit{Follow Path} & \cellcolor{bad} --- & \cellcolor{mid} \cite{Ware1996}, \cite{Ware2005}, \cite{Ware2008} & \cellcolor{bad} --- \\
        \cmidrule{2-5}
        & \textit{Revisit} & \cellcolor{bad} --- & \cellcolor{mid}\cite{Kotlarek2020} & \cellcolor{mid}\cite{Kwon2016} \\
        \cmidrule{2-5}
        & \textit{Group Only Tasks} & \cellcolor{inc}\textbf{Ours}, \cite{Alper2011} & \cellcolor{good}\textbf{Ours} & \cellcolor{good}\textbf{Ours} \\
        \cmidrule{2-5}
        & \textit{Group-Link Tasks} & \cellcolor{good}\textbf{Ours} & \cellcolor{good}\textbf{Ours} & \cellcolor{good}\textbf{Ours} \\
        \cmidrule{2-5}
    \end{tabular}

    \label{tab:task_layout_matrix}
\end{table*}

\subsection{Network Communities and Tasks}
There are many factors that can add complexity to a simple network. Nodes can have a set of attributes, and edges can be weighted and/or directed. All of these additional data can be used to infuse the network with more structure. A common practice is to group nodes based on shared attribute values. The result is a hierarchical structure---a community structure. Nodes are said to be part of the same community if they share similarities in some attribute. 
The leaf nodes in the hierarchy correspond to the nodes of the original network. 
The higher-level nodes in the hierarchy are called meta-nodes, which form groups by aggregating their lower levels.

Tasks performed on regular networks have been thoroughly categorized~\cite{Lee2006, amar2005low}. These taxonomies, however, do not concern themselves with community-based networks. Saket et al.~\cite{Saket2014} have compiled a taxonomy specific to community-related tasks. The tasks they describe are the ones commonly used to evaluate visualizations of networks with group structures. In other cases, it is important to identify communities in networks rather than operating on them. Fortunato et al.~\cite{fortunato2010community} give an exhaustive overview of this topic. Greffard et al.~\cite{Greffard2015} have investigated the potential of stereoscopy while Cordeil et al.~\cite{Cordeil2017} have compared different types of immersive technology for these types of tasks. Although this proves the potential of VR in that area, we limit ourselves to the type of tasks formerly described~\cite{Saket2014}, where communities are already established.

\subsection{Community Drawing}
Drawing networks with complex community structures poses a challenge. General networks are traditionally drawn using force-directed layouts~\cite{Kobourov2012}. These common methods often fail to communicate the community structure of a hierarchical network. There have been efforts to develop layouts to accommodate a hierarchical network~\cite{Vehlow2017}. Several studies, such as Johnson et al.~\cite{Johnson1991}, Bourqui et al.~\cite{Bourqui2007}, Itoh et al.~\cite{Itoh2009} or Didimo et al.~\cite{Didimo2014}, have introduced layout methods for the disjoint, hierarchical type of networks described by Vehlow et al. \cite{Vehlow2017}. The proposed methods emphasize the community structure by either leaving out some edges and trying to maximally utilize the available space~\cite{Johnson1991}, by using recursive space division to group similar nodes using multi-level force-directed layouts~\cite{Bourqui2007}, or by combining different methods to layout the graph and communities separately~\cite{Itoh2009}. Others have focused on improving performance of existing approaches~\cite{Didimo2014}. 

Another challenge in network visualization is handling a large number of edges. A common technique to overcome visual complexity is edge bundling. Edges will appear as bundles in parts where they share similar trajectories. A number of authors have developed ways to achieve this bundling effect~\cite{Phan2005Flow, Ersoy2011, Holten2006, Holten2009}. Holten~\cite{Holten2006} has proposed a method to apply edge bundling to hierarchical networks by taking advantage of common meta-nodes among leaf nodes that share the same attributes. We deal with networks that are inherently hierarchical, however rendering the network in 3D requires special considerations for occlusions, layout, and projections as shown by Kwon et al.~\cite{Kwon2016}. In this project we adapt some of the layout as well as bundling techniques~\cite{Kwon2016, Holten2006, Itoh2009, Didimo2014} to de-clutter the network in VR space.

\subsection{3D Network Visualization}
Visualizing abstract data in three dimensions bears many problems. Although there have been attempts of visualizing networks in 3D~\cite{Robertson1991, Rekimoto1993, Munzner1997, Young1996}, they invariably suffer from occlusion and the need for additional depth cues~\cite{herman2000graph} when displayed on two-dimensional viewing devices (such as a standard computer monitor or display wall). This is why the topic has not received as much attention in recent years. The depth cues available to 3D renderings projected to 2D space are limited to occlusions, motion parallax and shading~\cite{ware2019information, cutting1995perceiving}. However, the introduction of stereoscopic viewing and VR has brought new possibilities to this field. These new technologies allow users to freely move about and explore the data. Most importantly, immersive displays offer an additional depth cue: binocular disparity. McIntire et al.~\cite{McIntire2015} have shown that this can improve task performance in visualization tasks. Network visualizations are particularly suitable, as improved task performance was observed in multiple studies~\cite{Ware1996, Ware2005, Ware2008}. Aside from low-level tasks like the ones that the former authors investigated, community identification tasks have also been shown to benefit from stereoscopy~\cite{Greffard2015}.

Works in this area have since matured and explored more ways to handle networks in three-dimensional space. Kwon et al.~\cite{Kwon2016} have analyzed different projection methods for spherical layouts to maximize space utilization in VR. Büschel et al.~\cite{Bueschel2019} have compared the suitability of different edge styles in AR. Others have investigated possibilities of novel interaction design for those environments~\cite{Alper2011, Kotlarek2020, Huang2017, Halpin2008, Sorger2019}. In terms of interaction, our work is most closely related to the works of Alper et al.~\cite{Alper2011}, Kotlarek et al.~\cite{Kotlarek2020}, and Sorger et al.~\cite{Sorger2019}. Their proposed interaction techniques have inspired us to build on top of the ideas of overview and detail and stereoscopic highlighting. Huang et al.~\cite{Huang2017} have explored the feasibility of gesture controls in virtual reality. Different navigation interactions have been compared by Drogemueller et al.~\cite{Drogemueller2019}. Both the gestures and navigation paradigms lend themselves excellently for more complex interactions or for systems that feature a broad array of simultaneously available functions. 
\begin{figure*}[!htb]
  \centering
  \includegraphics[width=1\textwidth]{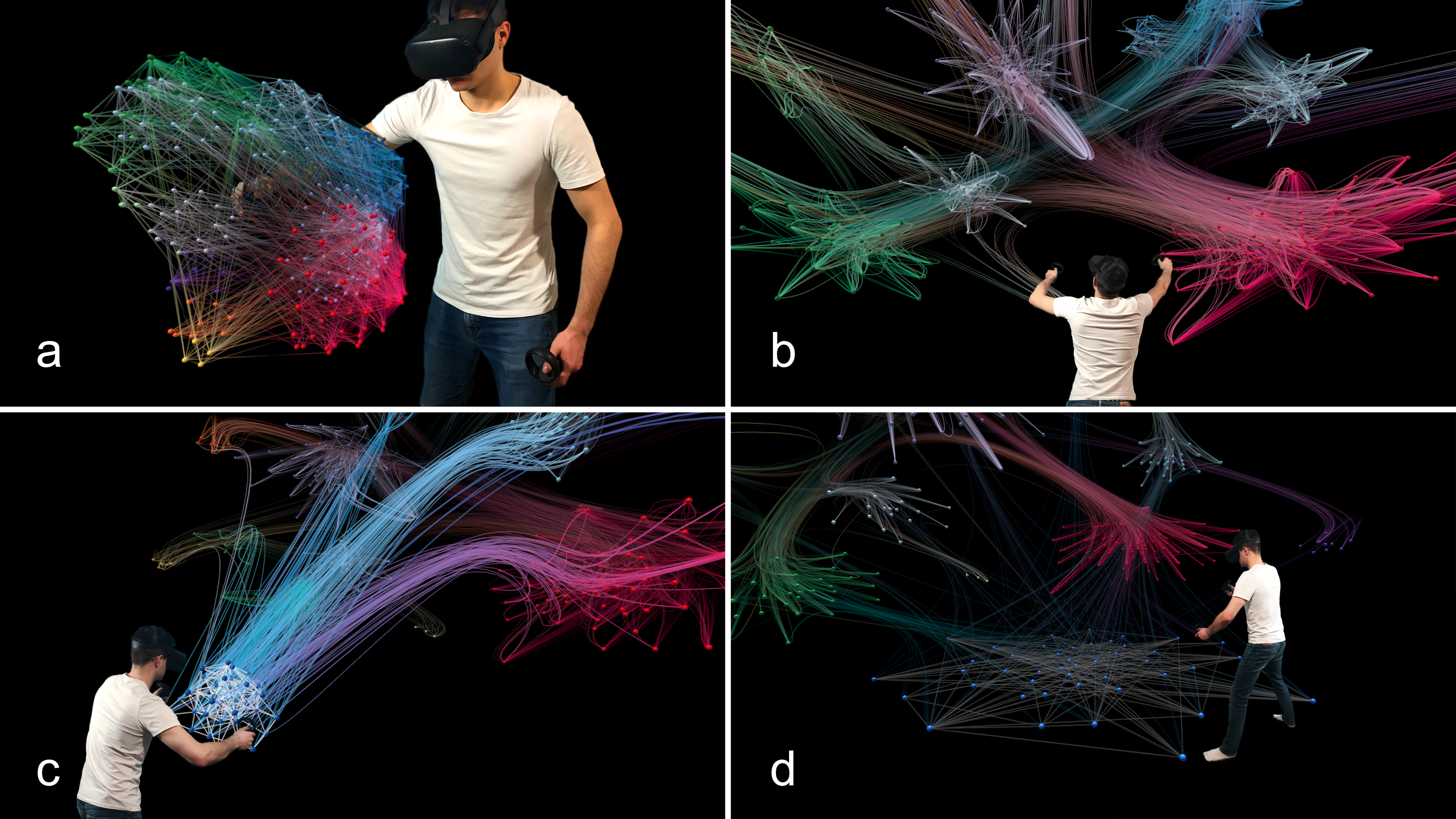}
    \caption{(a) The network is shown as a miniature floating in front of the user. The H3 hyperbolic layout~\cite{Munzner1997} is used to layout the nodes. (b) In the spherical view, the network is projected onto a large sphere. The view is scaled to fill the horizon and take up about 180 degrees of the user's field of view. (c) Selected communities descend from the spherical layout for the user's close inspection. The group expands, and the rendering and layout of the community changes to reveal more detail about intra-community connections while emphasizing its place in the context of the whole network. (d) Expanded communities can be projected to the floor to lay bare details of intra-community connectivity that might not be apparent in the three-dimensional view.}
    \vspace{-0.1in}
  \label{fig:multi_collage}
\end{figure*}

\section{Methods}
Virtual reality (VR) allows for intuitive exploration of 3D space. We leverage the additional depth cues to present the data to the user. However, no graph layout method is universally applicable for all possible tasks, even with these perceptual advantages. We present a multi-layout approach that lets users interactively adapt the layout of a hierarchical network and its communities as she sees fit for the task at hand. Specifically, we propose a suite of four layouts that maximize the utility of available space in VR. To the best of our knowledge and in contrast to prior work on immersive network visualization, our novel visualization design is the only one to support a broad variety of tasks and layouts (Table~\ref{tab:task_layout_matrix}).

\subsection{Multi-Layout Suite}
Our system comprises four views. Each view shows all the data but emphasizes different aspects of the network and is intended to support different tasks. Prior work is limited to single layouts or simple overview plus detail concepts, supporting only a limited number of tasks. With our approach, we hope to bridge the gaps and provide users with a system that facilitates efficient graph exploration and task completion. For our design, we picked four views that are orthogonal in their utility to that end. Each layout reveals a different aspect of the network, and the flow between them is organized by level of detail, ranging from a broad overview down to a layout that allows detailed inspection of single nodes and links. This workflow expands on the idea of overview plus detail by breaking it up into multiple levels. In addition to that, the choice for these four layouts and the drill-down workflow was informed by available task taxonomies~\cite{Lee2006,Saket2014} which suggest the need for attention on multiple levels of detail--especially for networks with group structures. With this suite of layouts, we expect users to be able to efficiently solve such tasks. Table~\ref{tab:task_layout_matrix} highlights the capabilities of our systems and terms of graph layouts and supported network tasks. It puts our contribution into perspective with prior work in network analysis.

To get a general impression of the data, a miniature overview of the network is placed in front of the user (Figure \ref{fig:multi_collage} (a)). After an initial inspection of the network, the user can expand this view into a larger spherical projection that stretches the horizon and makes optimal use of the available space (Figure \ref{fig:multi_collage} (b)). In this view, communities and their general connectivity become more apparent. The user might decide to explore one or more of them in more detail. Selecting communities will expand its nodes into a separate three-dimensional force-directed layout that descends to the user for closer inspection (Figure \ref{fig:multi_collage} (c)). Finally, this floating layout can be projected to the ground below the user, where it further expands into a flat 2D layout taking up most of the user's field of view (Figure \ref{fig:multi_collage} (d)). 

In conjunction, these layouts constitute a continuous workflow that takes the user from a high-level view of the whole network down to the level of single nodes within a community. In the following sections, we discuss each layout and their interactions in more detail. Each section is accompanied by a real-life use-case scenario based on one of our datasets~\cite{celegansneural}. The data represents the nervous system of a C. Elegans worm encoded as a network. Each node represents a neuron. Edges indicate functional connections between those neurons. We show the workflow of finding neurons with strong and diverse inter-community connectivity. For a visual tour through this workflow, please refer to the supplemental video.

\subsection{Overview Layout}
The overview is meant to convey a general impression of the structural properties of the hierarchical network (Figure \ref{fig:multi_collage} (a)). We use the H3 hyperbolic layout~\cite{Munzner1997} to calculate node positions. The layout was chosen as it is especially suited for visualizing large networks with inherent group structures which is the type of network we use in this project. Links are drawn as straight lines, and color is used to encode communities. The network is scaled down to fit the user's view comfortably and is placed in front of her at about shoulder's height in VR space. As this view is meant to introduce the viewer to the dataset, the placement and scale were chosen such that she can freely walk around the network. The experience is similar to looking at a museum art piece in a display case from all angles. At the touch of a button on the motion controller, the miniature expands into a larger spherical layout.

\textbf{\textsc{Use Case}} \textit{The user identifies the overall structure of the neural network at a glance. Color coding of communities allows her to quickly identify 13 major functional groups of the nervous system. She is also able to see that some groups are more densely connected than others.}

% \begin{figure}[!htb]
%   \centering
%   \includegraphics[width=1\columnwidth]{images/multi-layout/overview.png}
%     \caption{The network is shown as a miniature floating in front of the user. The H3 hyperbolic layout~\cite{Munzner1997} is used to layout the nodes.}
%   \label{fig:multi_overview}
% \end{figure}

\subsection{Spherical Layout}
Virtual reality environments offer a vast space that surrounds the user. This layout facilitates this space to the highest degree. Spherical projections have been shown to be a suitable method of visualizing a network in virtual reality environments~\cite{Kwon2016}. They use the available space by placing the user in the center of the projection and therefore providing an equidistant placement of elements in a network in relation to the user's point of view. 
%Depending on the type of projection, different attributes of the original 2D layout can be preserved (e.g., straight lines, the distance between points, etc.). 
Aside from this factor, the field of view (FOV) utilized by the projection can impact user performance and comfort. Note that we use the term FOV to mean the horizontal and vertical angular space taken up by the layout on the sphere. The FOV of the human eye (i.e., the angular range perceptible to us at any given point in time) is fixed for each individual. By varying the FOV, the user's level of immersion ranges from seeing the network just in front of oneself (low utilization of the sphere, low FOV) to being completely enveloped by elements of the network (full utilization of the sphere, high FOV).

% \begin{figure}[!h]
%   \centering
%   \includegraphics[width=0.5\linewidth]{images/SphericalLayout.png}
%   \caption{Network nodes are projected onto the surface of the sphere. The total FOV taken up by the nodes is about $180^{\circ}$. The exact FOV varies slightly on a per-layout basis.}
%   \Description{A network drawn in VR is shown. Nodes are arranged on the surface of a sphere. They take up approximately half of the spherical surface. For demonstration, the sphere is hinted at in semi-transparent white.}
%   \label{fig:SphericalLayout}
% \end{figure}

Limiting the FOV can reduce exaggerated head movement and increase task performance and comfort, as shown in previous studies~\cite{Kwon2016, Youdas1992}. Therefore, we decided to fixate the FOV to a value of just under $180^{\circ}$ both horizontally and vertically.

The spherical layout constitutes only part of the technicalities since it merely projects an exiting 2D layout onto a sphere. The base layout can also affect the outcome visually. The one we chose is derived from the works of Itoh et al.~\cite{Itoh2009} and Didimo et al.~\cite{Didimo2014}, who both investigated layout methods for hierarchically clustered networks. These networks are the same type of community-based networks that we are treating. Communities are groups of leaf nodes in a hierarchically clustered network. Nodes that share the same parent are said to be in the same community. 

To calculate the clustering, we use the Louvain method for community detection \cite{blondel2008fast}. For the layout, we employ much of what Itoh et al.~\cite{Itoh2009} and Didimo et al.~\cite{Didimo2014} have already proven to be effective. We first compute a node hierarchy with the aforementioned technique \cite{blondel2008fast} which is followed by a combination of a force-directed layout \cite{forceatlas2} for the group level and a treemap layout for the overall placement of communities. The resulting layout spans the horizon around the user (Figure \ref{fig:multi_collage} (b)) and edge bundling visually emphasizes inter-community connection patterns.

% \begin{figure}[!htb]
%   \centering
%   \includegraphics[width=1\columnwidth]{images/multi-layout/spherical.png}
%     \caption{In the spherical view, the network is projected onto a large sphere. The view is scaled to fill the horizon and take up about 180 degrees of the user's field of view.}
%   \label{fig:multi_spherical}
% \end{figure}

\textbf{\textsc{Use Case}} \textit{After gaining an overview of the data, the user expands the layout into the spherical view. She identifies several candidate groups that stand out due to their strong bundles of edges. These groups might play a central role in the worm's nervous system.}

\subsection{Floating Community Layout}
In its initial state after expansion, the network and all of its communities are subject to the spherical layout described above. Upon selecting a community of interest, its layout changes from the treemap-based hybrid technique to a separate, three-dimensional, force-directed layout (Figures \ref{fig:teaser}, \ref{fig:multi_collage} (c), \ref{fig:Interaction} (b), (d), (e)). In contrast to the overall network structure, the selected community's layout is calculated and updated continuously as it descends from the spherical projection closer towards the user. Dynamically updating the layout adds aesthetic value through fluidity in motion and smooth transitions from the group's initial state to its final constellation. Aside from aesthetics, it offers a sense of continuity. While the new layout unravels, the user sees the nodes and their connections in different constellations. This aids the user's understanding of the intra-group structure. Furthermore, the user can adjust the layouts arrangement---moving nodes around to get a better view of parts of the community that might be occluded. Links within the community change their style to straight edges and are rendered in clear white color. Similarly, links that connect outward to other communities are emphasized too. Connections unrelated to that particular group are subdued.

\textbf{\textsc{Use Case}} \textit{In this view, it is easy to discern neurons with strong external connectivity. By expanding multiple groups simultaneously, the user can find nodes that share a connection, make judgements about the interplay of neuron clusters, and identify communities that connect to a large number of other clusters.}

% \begin{figure}[!htb]
%   \centering
%   \includegraphics[width=1\columnwidth]{images/multi-layout/detail.png}
%     \caption{Selected communities descend down from the spherical layout for the user's close inspection. The group expands and rendering and layout of the community changes to reveal more detail about intra-community connections while emphasizing its place in the context of the whole network.}
%   \label{fig:multi_floating}
% \end{figure}

\subsection{Projected Community Layout}
While inspecting a floating community, the user might want more information about the connections within a group. Depth cues available in VR can help understand a community's structure, but there are cases where a traditional 2D projection reveals additional details. For such cases, the user can project the group of nodes down to the ground below her. The community changes to a flat, horizontally arranged, force-directed layout \cite{forceatlas2} that fills the space below the user's position (Figure \ref{fig:multi_collage} (d)). It can be viewed comfortably by looking down. Similar to the floating layout, intra-community links are pronounced. Inter-community links and unrelated connections still connect to the spherical layout above but are both greatly subdued, enabling the user to focus completely on the single community.

\textbf{\textsc{Use Case}} \textit{The user invokes the projected layout for a strongly connected community to spread it out across the floor. Here she has enough space to closely inspect all the nodes one by one. She identifies several nodes that have strong inter-community connectivity and records their names.}

% \begin{figure}[!htb]
%   \centering
%   \includegraphics[width=1\columnwidth]{images/multi-layout/throw-down.png}
%     \caption{Expanded communities can be projected to the floor to lay bare details of intra-community connectivity that might not be apparent in the three-dimensional view.}
%   \label{fig:multi_projected}
% \end{figure}

\subsection{Interaction}
\begin{figure}[!htb]
  \centering
  \includegraphics[width=1\columnwidth]{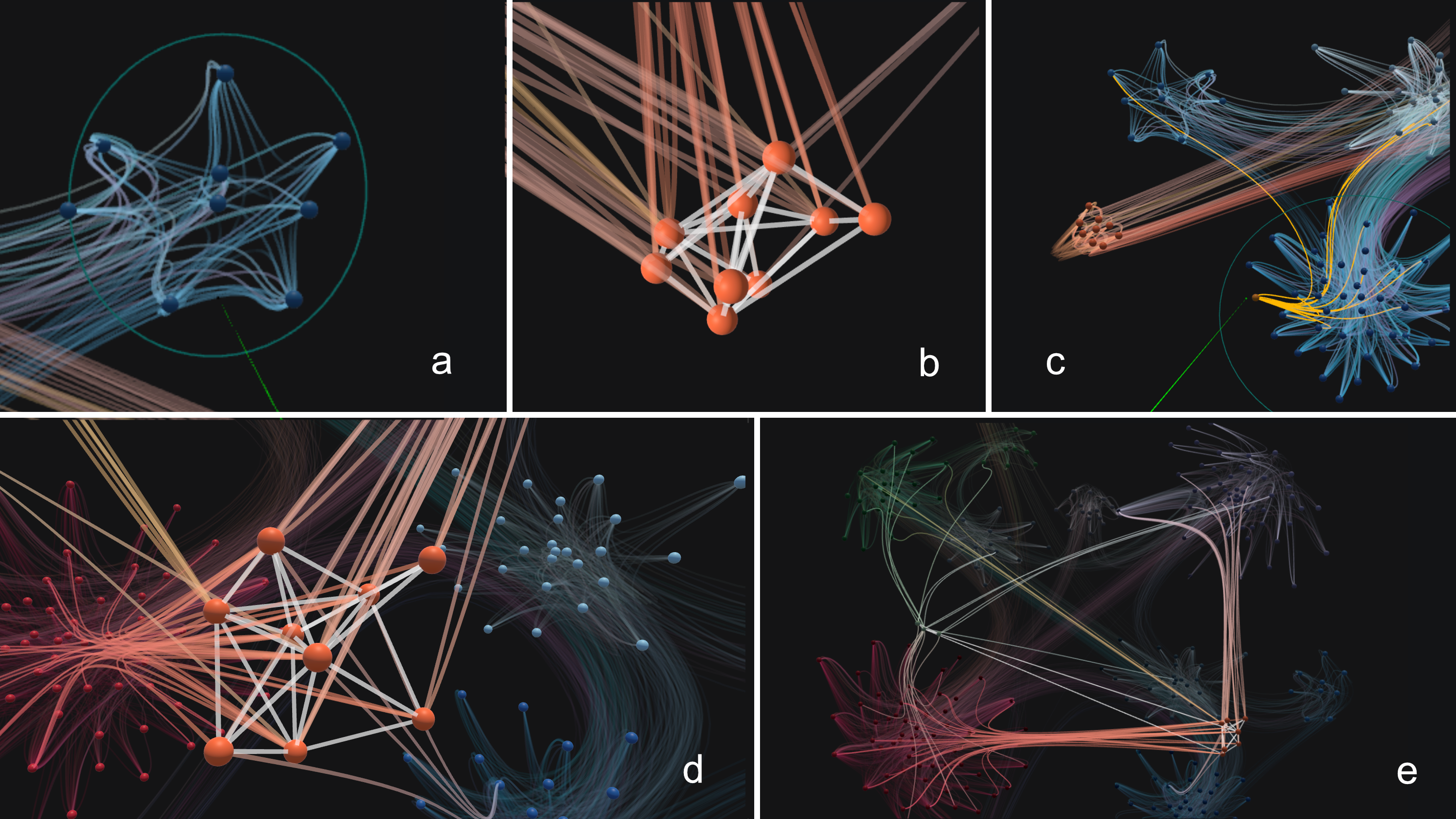}
    \caption{Some of the interaction techniques in our system. (a) Highlight Community. (b) Expand Community. (c) Highlight Nodes. (d) Effect of expanding a single community. Links between communities are emphasized while the rest of the network is subdued. (e) Effect of expanding multiple communities. Similar effect to (d). Links between expanded communities are rendered straight.}
    %\vspace{-0.5in}
  \label{fig:Interaction}
\end{figure}

Although interaction design is not a focus of this work, it was important for us to aim for high usability. Part of our design goals was to provide a set of interactions that are easy to execute yet effective at aiding the user in exploring the visualization. We decided to use the motion controls provided by the HTC Vive~\cite{VIVETM}. These controls allow developers to implement natural interactions like grabbing, pushing or pulling. In this work we deliberately decided against providing users with such gesture-based interactions. On the one hand, this was done to keep required training to a minimum. All the interactions in our system can be triggered by pressing a small set of buttons on the controllers. The other main reason for this decision was that we wanted the results of our study to be comparable to previous network visualization work in VR~\cite{Alper2011,Kotlarek2020,Sorger2019,Kwon2016}, where inputs were also mostly realized with simple button presses. In the system, a ray is cast, originating at the controller's position and extending towards where the user points with the controller. Previous studies have made successful use of similar techniques to enable interaction in immersive network visualizations~\cite{Kotlarek2020, Huang2017}. They enable users to point at and select objects, which is the main focus of our design. Much of the user's interaction centers around selecting and inspecting nodes, communities, and their connectedness. With these things taken into consideration, we offer the following interactions:

\begin{itemize}
\setlength{\itemsep}{1pt} \setlength{\parskip}{0pt} 
    \item \textbf{\textsc{Expand Network}} - Transitioning from the overview to the spherical layout (Figure \ref{fig:multi_collage} (a), (b)).
    \item \textbf{\textsc{Highlight Community}} - Pointing at a community with the ray. The community will be highlighted by a colored circle (Figure \ref{fig:Interaction} (a)).
    \item \textbf{\textsc{Expand Community}} - Expanding a highlighted community by pressing a button. The community will descend towards the user's position stopping right in front of her (Figures \ref{fig:teaser}, \ref{fig:multi_collage} (c), \ref{fig:Interaction} (b), (d), (e)). 
    %Nodes in the community will expand. Their layout will change from a spherical representation on the surface of the sphere to a true three-dimensional layout, and the rendering style of intra-community edges will change to straight lines instead of curved splines (Figures \ref{fig:Interaction} (b), (d) and (e), \ref{fig:multi_collage} (c)).
    \item \textbf{\textsc{Project Community}} - Projecting an expanded community will change it to a flat force-directed layout. The layout descends beneath the user taking up the available space there (Figure \ref{fig:multi_collage} (d)). %Intra-community edges will stay straight lines while their colors become more subdued. Other edges in the network will be almost transparent to shift the focus to the currently projected community (Figure \ref{fig:multi_collage} (d)).
    \item \textbf{\textsc{Reset Community}} - Resetting an expanded community back to its condensed state. The community will float back to its original position. Nodes in the community will contract, and the rendering style of intra-community edges will revert back to curved splines.
    \item \textbf{\textsc{Highlight Node}} - Nodes can be highlighted much like the way communities are highlighted. The node and all of its edges will change color to emphasize its relation to other nodes and communities (Figure \ref{fig:Interaction} (c)). This applies to all layouts so that highlighted links will always stand out even if their neighbors are heavily subdued.
\end{itemize}

\subsection{Edge Bundling}
The large number of edges that we display in our system and, more-so, the unique community structure that these networks exhibit demand special considerations for rendering. Hierarchical edge bundling has been successfully used in prior work to effectively improve the legibility of hierarchical networks~\cite{Holten2006}. Recently, these techniques have been expanded to accommodate bundles in spherical layouts~\cite{Kwon2016}. 

Our approach to bundling is inspired by previous work \cite{Kwon2016, Holten2006}. However, these methods do not satisfy the need to distinguish between intra-community edges and inter-community edges. In our design, edges between nodes in the same community are bundled locally. On the other hand, edges that connect nodes in two different communities are rendered using the previously proposed hierarchical edge bundling techniques~\cite{Holten2006, Kwon2016}.
We vary bundling strength depending on the currently active layout. Bundling strength ranges from $0.0$ to $1.0$. In the overview, edge bundling is disabled ($0.0$ bundling strength). The spherical view uses moderate bundling strength ($0.7$) to reduce visual clutter and emphasize overall connectedness between communities. When communities are expanded, their intra-community links are rendered as straight-line edges since their number is limited enough to still provide a good overview \cite{Yoghourdjian2018}. Other links in the remaining graph stay bundled ($0.7$ bundling strength). Upon invoking the projected community layout, bundling strength in the main graph is increased ($0.9$ bundling strength) to remove clutter further and help shift the focus to the projected group of nodes. In this view, intra-community links are again rendered as straight lines.

% \begin{figure}[!h]
%   \centering
%   \includegraphics[width=1\columnwidth]{images/Bundling.png}
%     \caption{Edge bundling strength ranging from $0.0$ (no bundling) to $1.0$ (full bundling).}
%     \vspace{-0.2in}
%   \label{fig:Bundling}
% \end{figure}

\subsection{Rendering}
For rendering nodes, we use GPU instancing. This can effectively reduce the number of draw calls by magnitudes. Edge positions and subsequent spline interpolations are calculated using compute shaders. Edge drawing is done by procedural draw calls available through Unity3D. The compute shader results are shared in GPU memory and are therefore available for drawing without having to transfer the data back to the CPU. To improve performance further, all edges are rendered in a single draw call. We use a regular shader to access compute shader results and generate edge geometry for the entire scene. To reduce the number of vertices, we employ billboarding so that each line segment requires only two triangles.

\begin{figure}[!h]
  \centering
  \includegraphics[width=1\columnwidth]{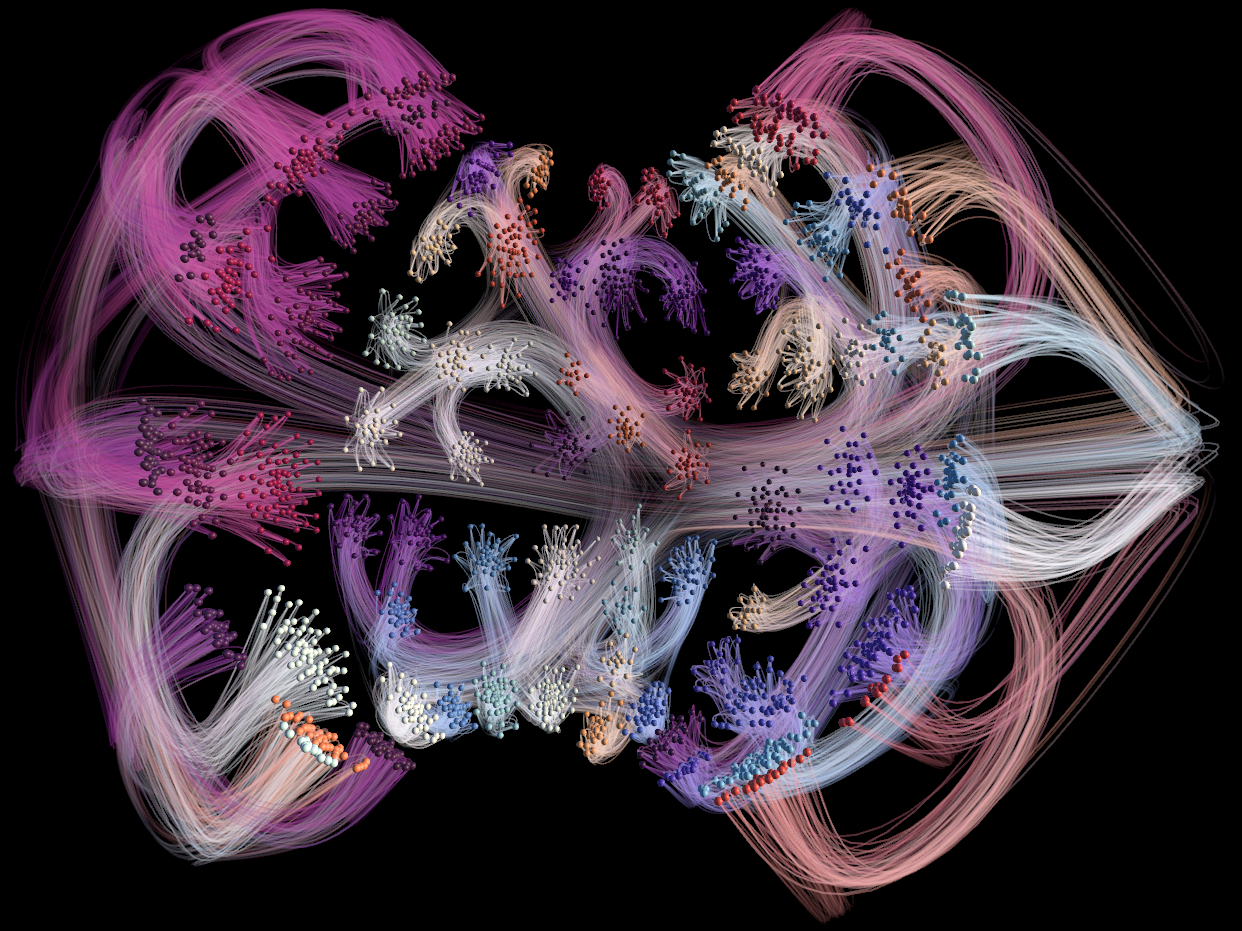}
    \caption{The largest network in our tests. It contains 2646 nodes and 10455 edges.}
  \label{fig:Rendering}
\end{figure}

The largest network in our tests consisted of 2646 nodes and 10455 edges (Figure \ref{fig:Rendering}). We achieved real-time frame-rates in VR while providing full interactivity.
%\subfile{sections/usecases}
\section{User Study}
The purpose of this study is to investigate the utility of our multi-layout approach to immersive visualization of network data. Previous studies have already shown that people tend to perform better in graph tasks when working in immersive environments as compared to 2D or even monoscopic 3D environments~\cite{Ware1996,Ware2005,Ware2008,Kwon2016,Kotlarek2020}. However, these studies only consider single, static graph layouts with minimal interactivity. In this study we examine the effects of multiple, interactively connected layouts to perform such graph tasks in immersive environments.

\subsection{Study Design}
We conducted a study with 20 participants to investigate how our multi-layout system performs against the state of the art of immersive network visualizations~\cite{Kwon2016}. To construct a baseline and ensure comparability, we chose a subset of features from our system that most closely represents the functionality of prior work~\cite{Kwon2016}. The experiment was designed as a between-subject study that had participants complete several network-related tasks. There were two conditions in total. Each participant completed \textit{1 condition $\times$ 3 networks $\times$ 5 tasks}. Networks were presented in ascending order of complexity. The sequence of tasks was randomized among subjects. We recorded accuracy, time to completion, total number of interactions, and separate counts for each type of interaction. 

\subsubsection{Visualization Conditions}
For this study, we considered two conditions:

\begin{itemize}
\setlength{\itemsep}{1pt} \setlength{\parskip}{0pt} 
    \item \textbf{MULTI} - Multi-layout suite in virtual reality. Displayed on a stereoscopic head-mounted display.
    \item \textbf{BASE} - State-of-the-art in immersive network visualization~\cite{Kwon2016} which offers a treemap-based spherical layout for the whole network and a detail view of individual groups.
\end{itemize}

\noindent For both conditions, we used the same basic settings in our system. These includes values like edge bundling strength, edge opacity, or animation speed within and between different layouts. 

The \textit{MULTI} condition offers all the features described earlier while the \textit{BASE} condition is modeled after Kwon et al.~\cite{Kwon2016} who offered a treemap-based spherical layout in which single groups could be brought closer to the user by selecting them. Our \textit{BASE} condition offers this same set of capabilities.

Both conditions use the same underlying hybrid layout technique for the spherical view, which is a combination of a treemap-based layout for the placement of communities and a force-directed approach \cite{forceatlas2} on the group-level. The floating layout also uses force-directed methods, each appropriate to the spatial dimensions they are shown in.

Some tasks require us to visually highlight nodes or groups of interest. Such targets are marked by in a distinct, colored indicator in both conditions. In tasks involved communities, a red circle is placed around those groups. For tasks that require the attention to single nodes, we use a red fill color which is easily distinguishable from regular colors used in the rendering. In addition, nodes are also emphasized through a small halo of that same color around their perimeter.

\subsubsection{Network Data}
We used three different networks for the study with an additional fourth network for initial training sessions. The networks are shown in ascending order of complexity. For the purpose of this study, we label these different levels of complexity as easy, medium, and hard. The network used for training is equivalent to the easy graph in terms of complexity. The exact metrics are shown in Table \ref{tab:data}.

\begin{table}[!htb]
    \centering
    \caption{Network complexities in terms of number of nodes, number of edges, number of communities, and the maximum hierarchical depth of the graphs used in this study.}
    \begin{tabular}{@{}lcccc@{}}
        \toprule
         & Nodes & Edges & Communities & H. Depth\\
         \midrule
         \midrule
        \textbf{Easy} & 115 & 613 & 12 & 2\\
        \textbf{Medium} & 297 & 2148 & 13 & 3\\
        \textbf{Hard} & 1133 & 5451 & 25 & 3\\
        \bottomrule
    \end{tabular}
    \label{tab:data}
\end{table}

\subsubsection{Tasks}
Our tasks are based on taxonomies proposed earlier~\cite{Saket2014, Lee2006, amar2005low} to maximize comparability among similar studies. 
In our study, we categorize them based on the domain the user has to act upon. We therefore define the following task categories:

\begin{itemize}
\setlength{\itemsep}{1pt} \setlength{\parskip}{0pt} 
    \item \textbf{\textsc{Intra-Community}} - Task is performed in the context of a single community. Based on Lee et al.~\cite{Lee2006}, Amar et al.~\cite{amar2005low}, and Saket et al.~\cite{Saket2014}.
    \item \textbf{\textsc{Inter-Community}} - Task has user focused on elements in multiple communities. Based on Saket et al.~\cite{Saket2014}
    \item \textbf{\textsc{Epi-Community}} - Task performed by looking at the whole population of communities. Based on Saket et al.~\cite{Saket2014}
\end{itemize}

Building upon the established taxonomies~\cite{Saket2014, Lee2006, amar2005low} and our newly introduced categories we, chose the following five tasks for this study.

\begin{itemize}
\setlength{\itemsep}{1pt} \setlength{\parskip}{0pt} 
    \item \textbf{\textsc{CommonNeighbors}} - \textit{Find Common Neighbors.} In a given community, find the nodes that are shared neighbors between two nodes. (\textit{intra-community})
    \item \textbf{\textsc{CountNodes}} - \textit{Count Nodes in a Community.} Given a community, count the number of nodes in it. (\textit{intra-community})
    \item \textbf{\textsc{MaxDegreeNode}} - \textit{Find Maximum Degree Node.} Given four nodes in the network, find the one with the highest degree. (\textit{inter-community})
    \item \textbf{\textsc{MaxDegreeCommunity}} - \textit{Find the Maximum Degree Community.} Among all communities, find the one with the highest number of outbound links. (\textit{inter-community})
    \item \textbf{\textsc{CountCommunities}} - \textit{Count number of neighboring Communities.} Given a community, count the number neighboring communities. (\textit{epi-community})
\end{itemize}

Our choice of tasks was informed by two considerations. The majority of tasks should be specific to networks with group structures. Such tasks were adopted from the domain-specific task taxonomies mentioned above~\cite{Saket2014}. In addition to that, we wanted to establish a link to prior work by including some tasks from Kwon et al.'s study~\cite{Kwon2016} for the purpose of comparability.

\subsection{Hypotheses}
Based on our task categories and related work, we defined the following hypotheses:

\begin{itemize}
\setlength{\itemsep}{1pt} \setlength{\parskip}{0pt} 
    \item \textbf{H1} - Our \textit{MULTI} design outperforms the \textit{BASE} version in \textit{intra-community} tasks.
    \item \textbf{H2} - Our \textit{MULTI} design does not outperform the \textit{BASE} version in \textit{inter-community} tasks.
    \item \textbf{H3} - Our \textit{MULTI} design outperforms the \textit{BASE} version in \textit{epi-community} tasks.
\end{itemize}

We justify our choice for \textit{H1} and \textit{H3} based on the affordances that our layout-suite offers. The projected community layout allows users to gain more detailed insights into intra-community constellations, while the overview layout offers an at-a-glance view of the data, which we expect to be conducive to solving such tasks. For \textit{H2}, we suspect that the split attention that is required to solve \textit{inter-community} tasks outweighs the benefits of having multiple views of the data~\cite{MillerMagic1956,Sweller1998,Kotlarek2020}.

\subsection{Apparatus}
Both conditions were implemented in Unity3D. We used a HTC Vive \cite{VIVETM} head-mounted display device (HMD) with both of the included motion controllers. The application was rendered on an NVIDIA GeForce RTX 2080 TI graphics processor and maintained a smooth framerate of about 78 frames per second. Participants were standing in both conditions to allow for more freedom of movement. The space available for them to move around in was about 60 square feet.

\subsection{Participants}
For the study, we recruited 20 participants (12 female, 8 male). Ages ranged from 18 to 34. None of the participants were colorblind. Eight wore glasses, and three wore contacts. The other nine participants reported no visual impairments. The design of the HMD allowed participants to leave on their glasses when wearing it. We adjusted the fit and ocular distance of the HMD for each participant to meet individual needs.

Four rated themselves as experienced with network data, nine reported moderate knowledge, six had little knowledge and one had never worked with graphs before. As for VR, three participants were experienced users, while eight had at least tried VR before. Nine participants had never used this technology, of which one had never heard of it.

\subsection{Procedure}
The group of participants was randomly divided into two equally-sized subgroups of which one performed the study in the \textit{BASE} condition while the other subgroup used the \textit{MULTI} version. Before the start of the study, each person was informed about the possible adverse effects of using a head-mounted VR device. These include cybersickness, vertigo, eye strain, or disorientation. They were told that they could pause or exit the study at any point in time if they felt uncomfortable for any reason. Due to recent public health events, all equipment involved in the study was thoroughly disinfected between each use, proper social distancing was practiced and face masks were worn at all times.

\subsubsection{Questionnaire}
Participants were asked to fill out a preliminary questionnaire on demographics and prior experience with networks and virtual reality.
In-between tasks, participants were required to answer questions about their experience through a set of NASA TLX \cite{hart1988NASATLX} questions. After the completion of all the tasks, participants answered several questions to summarize their experience in the experiment. This also gave them the opportunity to provide some unstructured feedback.

\subsubsection{Protocol}
Each participant was allowed to verbalize their actions and thoughts during the completion of tasks. This was limited to a degree that would not impact the user's performance. We created audio recordings to capture the dialog between the conductor and participants and "think-aloud" events during the experiment. 

\subsubsection{Training}
Before engaging in any real tasks, participants were instructed to complete a short training. It had users perform the same types of tasks found in the real experiment on a training network. This network was less complex than most of the ones used in the experiment and served to familiarize the participants with the nature of the tasks, the interactions and the layouts used in each condition. They were able to ask questions to clarify any issues and get feedback about the correctness of each task.

\subsubsection{Experiment}
In the experiment, participants were asked to complete the same kind of tasks that they knew from the training. The tasks were performed on three different networks. The networks were given in ascending order of complexity. The task order was randomized between subjects. No feedback about the correctness of the participant's answers was given. Participants were allowed to ask the conductor for information from the training slides. After the completion of each task category, participants were asked to provide feedback concerning this specific type of task. Responses were provided verbally and recorded by the experiment conductor as to not disrupt immersion and save time on the subsequent execution of further tasks. 
\section{Results}
Null hypothesis testing (NHT) has been under some criticism in different fields due to its limitations in reporting evaluation results \cite{AntiNHST, AntiNHST2}. We therefore provide confidence intervals (CI) and effect sizes in addition to traditional NHT results. Cohen's $d$ was used to calculate effect sizes. Quantitative results, for the most part, pass normality tests and, considering the relatively low sample count, are assumed to be normal distributed. CIs were taken from the respective t-distribution.

Overall, our results show that \textit{MULTI} outperforms \textit{BASE} in terms of accuracy for the majority of tasks. Time to completion was comparable in both conditions for most tasks. This means that users were able to perform tasks more accurately in the same time when using our multi-layout approach. Both \textit{CommonNeighbors} and \textit{CountNodes} were significantly better in the \textit{MULTI} condition, confirming H1. The inter-community tasks did not show any significant difference, which confirms H2. Lastly, H3 is confirmed by the significantly better performance in \textit{CountCommunities} for \textit{MULTI}. User feedback indicates a general preference for \textit{MULTI} with criticism mostly directed towards the positioning of graphs in space. In this section, we highlight the most important findings. For more data and analyses, please refer to the supplemental material.

\subsection{Intra-Community Tasks}
Intra-community tasks (Figure \ref{fig:ci_tasks} (a), (b)) are locally focused and do not require users to adjust their view much.
We saw significantly better overall accuracy in the \textit{MULTI} condition for \textit{CommonNeighbors} (Cohen's $d=1.467$, $p<0.01$) and \textit{CountNodes} (Cohen's $d=0.659$, $p<0.05$). For \textit{CommonNeighbors}, overall accuracy in \textit{BASE} was $0.478$ ($SD=0.275$) and $0.878$ ($SD=0.262$) for \textit{MULTI} while \textit{CountNodes} results show $0.869$ ($SD=0.173$) for \textit{BASE} and $0.957$ ($SD=0.069$) for \textit{MULTI}. Time to completion was not significantly different in either condition with Cohen's $d=-0.074$, $p=0.775$ and $d=0.134$, $p=0.605$ and with averages of $136.821$ ($SD=105.114$) and $68.156$ ($SD=48.198$) for \textit{BASE} and $129.789$ ($SD=79.94$) and $75.221$ ($SD=54.934$) for \textit{MULTI} for \textit{CommonNeighbors} and \textit{CountNodes} respectively.

\subsection{Inter-Community Tasks}
These tasks (Figure \ref{fig:ci_tasks} (c), (d)) require users to look around and adjust their view more frequently. For this category, we saw no significant difference in accuracy in either of the tasks. For \textit{MaxDegreeNode}, Cohen's was $d=0.155$, $p=0.549$ and averages were $0.734$ ($SD=0.442$) for \textit{BASE} and $0.8$ ($SD=0.4$) for \textit{MULTI}. \textit{MaxDegreeCommunity} had a Cohen's $d=0.382$, $p=0.145$ and averages of $0.433$ ($SD=0.496$) for \textit{BASE} and $0.622$ ($SD=0.477$) for \textit{MULTI}. Completion times for \textit{MaxDegreeNode} were significantly higher in \textit{MULTI} (Cohen's $d=1.339$, $p<0.01$) with averages of $65.292$ ($SD=38.204$) and $136.325$ ($SD=63.089$) for \textit{BASE} and \textit{MULTI}. There was no difference in completion time for the \textit{MaxDegreeCommunity} tasks with Cohen's $d=0.051$, $p=0.843$ and averages of $99.733$ ($SD=113.098$) and $104.7$ ($SD=72.932$) for \textit{BASE} and \textit{MULTI}.

\subsection{Epi-Community Tasks}
\begin{figure*}[t]
\centering
  \includegraphics[width=1\textwidth]{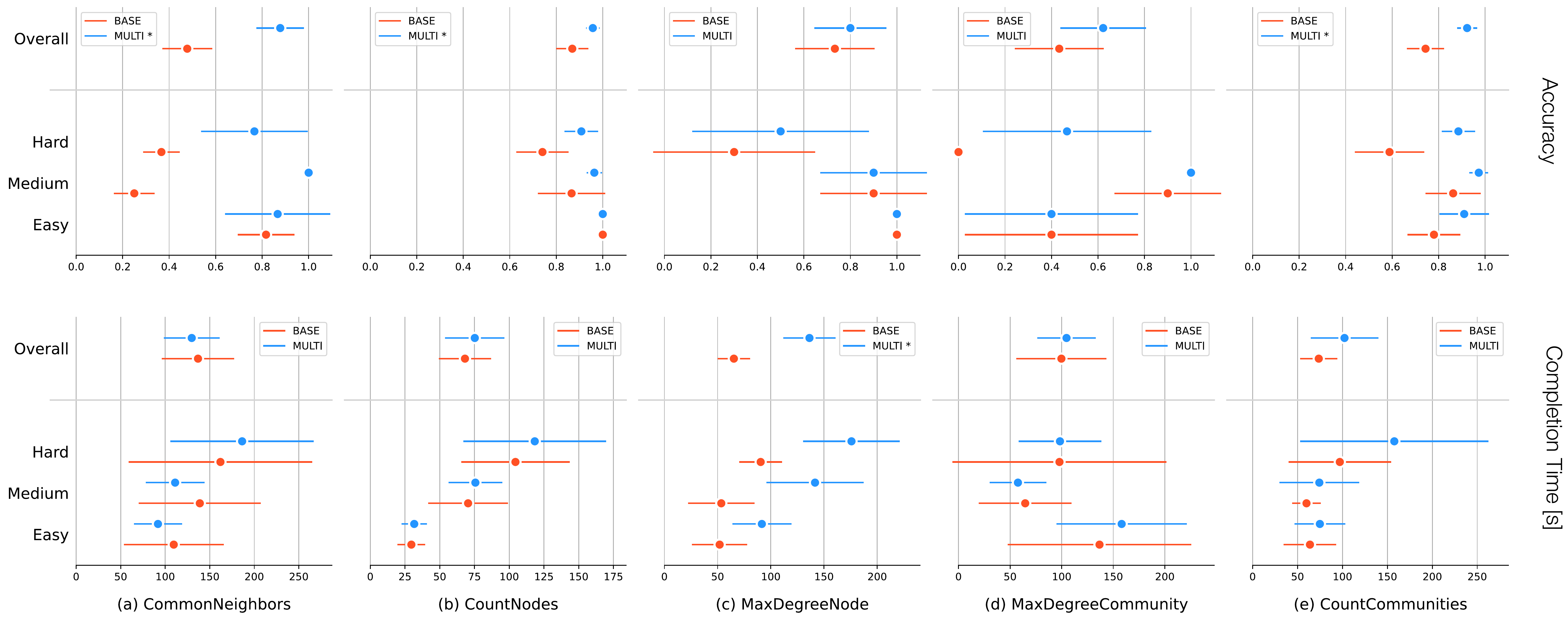}
  \caption{Performance results for all tasks given as the mean over all participants. The error bars denote 95\% CIs. Tasks are aligned in columns ranging from intra-community (a) and (b), over inter-community (c) and (d) to epi-community (e). The top row shows values for accuracy while the bottom lists task completion time in seconds. A star in the legend marks a significant difference in the overall results.}
  \label{fig:ci_tasks}
\end{figure*}

Participants need to inspect the whole network in this task (Figure \ref{fig:ci_tasks} (e)). Overall accuracy was significantly higher in \textit{MULTI} (Cohen's $d=1.091$, $p<0.01$) with averages of $0.744$ ($SD=0.202$) for \textit{BASE} and $0.923$ ($SD=0.106$) for \textit{MULTI}. Completion times were comparable in both conditions (Cohen's $d=0.362$, $p=0.167$). The overall averages were $73.619$ ($SD=52.338$) and $102.272$ ($SD=96.984$) for \textit{BASE} and \textit{MULTI}.

\begin{figure*}[t]
\centering
  \includegraphics[width=1\textwidth]{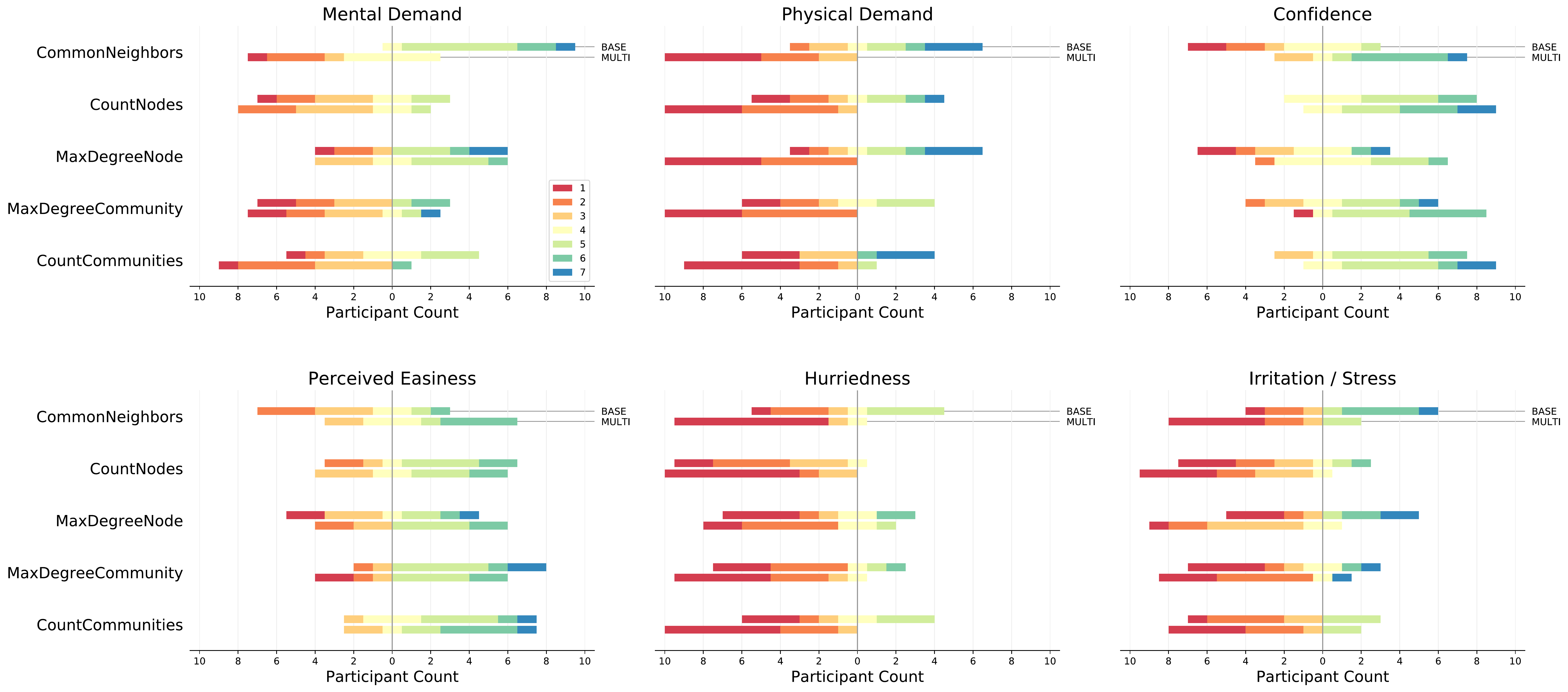}
  \caption{User's assessment of the tasks. Each graph shows one item of the NASA TLX~\cite{hart1988NASATLX} score with two bars for each task. Top bars are the \textit{BASE} condition, bottom bars are \textit{MULTI}. Color encodes selected value from red (values 1-3) to blue (values 5-7). Bars are centered around neutral ratings (4). Length denotes the frequency of that score in the assessment.}
  \label{fig:ci_feedback}
\end{figure*}

\subsection{User Feedback}
Participants were asked to provide unstructured feedback at the end of the study. 

For the individual tasks, users were asked several questions related to user experience. These questions were a subset of the NASA TLX~\cite{hart1988NASATLX} evaluation scheme. They were presented on a 1 to 7 Likert scale in which negative ratings were indicated towards 1 (e.g. not easy, low confidence, etc.), and positive ratings were on the 7 side (e.g. very easy, high confidence, etc.). These results are shown in Figure~\ref{fig:ci_feedback}. More detailed analysis, including CI and effect sizes in addition to NHT, similar to the task results, can be found in the supplemental material. 

Overall, users felt more confident (Cohen's $d= 0.687$, $p<0.05$) about their answers using the \textit{MULTI} condition while perceiving less mental (Cohen's $d=-0.437$, $p<0.05$) and physical demand (Cohen's $d=-1.503$, $p<0.01$). While perceived easiness (Cohen's $d=0.130$, $p=0.517$) of the tasks are the same in both conditions, users also felt less hurried (Cohen's $d=-0.800$, $p<0.01$) and less irritated (Cohen's $d=-0.596$, $p<0.01$) using our system.

Participants noted some criticism of the system:
\begin{itemize}
\setlength{\itemsep}{1pt} \setlength{\parskip}{0pt} 
    \item "Pulling groups closer felt too close."
    \item "My map was always located above me, so I had to constantly be looking at the ceiling."
\end{itemize}

The \textit{MULTI} condition also received several positive comments:
\begin{itemize}
\setlength{\itemsep}{1pt} \setlength{\parskip}{0pt} 
    \item "Miniature graph overview is very helpful for tasks that require a rough estimate."
    \item "The projection [layout] is good to find common neighbors of two nodes [...]"
    \item "[...] the function where you could expand the nodes on the "floor" [...] was very helpful in some of the tasks."
\end{itemize}

% \biblio
% \end{document}
% \documentclass[../paper.tex]{subfiles}

% \begin{document}

\section{Discussion}
The results of our study confirm our hypotheses and hint towards several trends. Users were more accurate in spatially focused tasks (\textit{intra-community}) using the \textit{MULTI} condition while taking the same time as the baseline. These tasks primarily concerned single nodes or nodes within the same community. Due to this, users in both conditions tended to expand the community in question. In the multi-layout approach, most expansions were followed by a projection to see even more details. This underlines the utility of the projected layout.

On a larger scale---for tasks that focus on multiple communities (\textit{inter-community})---users in the \textit{MULTI} condition did not perform better than the baseline. This is reflected in the interaction statistics. Aside from community expansion, which helped users track external links from expanded communities back to their source in the main network more easily, the remaining layouts were used very little. Interestingly, users took significantly longer to solve the \textit{MaxDegreeNode} task in the \textit{MULTI} condition. We suspect this to be due to the relatively higher utilization of the projected layout in this task. Users tended to spend a long time inspecting nodes but were ultimately unable to perform better than the baseline. These results indicate that the additional views on the data were not helpful to solving the tasks. We attribute this to the fact that the spherical view balances overview and detail quite well, which reduces the amount of split attention required to solve the task and can therefore be beneficial when estimating node or group degrees. Due to the lack of layout utilization, the equality in performance in both conditions is not surprising. 

Lastly, the \textit{epi-community} task required the users to focus on the whole network. Our multi-layout approach again outperformed the baseline for this type of task. The projected view was utilized more often than the overview, which is surprising since we initially suspected the overview to be more useful here. Our data indicates that the projected view helped users distribute a large number of inter-community links more broadly, making the counting task more manageable. Contrary to our expectations, users in the \textit{BASE} condition made almost no use of the expansion feature despite having experienced its relative benefits when compared to the spherical view.

To summarize, our results and user feedback indicate that the larger the network and the more spatially complex the task, the better the results can be expected from the multi-layout approach. Accuracy varied among tasks and conditions, while average completion times were generally comparable. These findings align with observations made by Greffard et al. \cite{Greffard2015} in that users tend to take the same time to complete tasks while approaching them differently in terms of focus and interactions. Unsurprisingly, the narrower the scope of a task, the less frequent the different layouts were used. As tasks increased in complexity and scope, layouts tended to be switched more often on average. Overall, the number of interactions varied greatly between users, hinting at vastly different problem-solving approaches. Ultimately, the tools we offered in our method enabled users to perform significantly better in most tasks which shows that a multi-layout approach performs well in a wide range of tasks. These results are encouraging insofar as they confirm prior observations about the viability of immersive visualization for network analysis~\cite{Ware1996, Ware2005, Ware2008, Kwon2016, Kotlarek2020}. Moreover, they show that by carefully designing layout and interaction suits for VR, we are able to build analysis systems that can greatly improve users' productivity. These results are proof that we have not reached the full potential of what immersive visualization has to offer. Nonetheless, through our study, we show the superior performance of an immersive multi-layout suite over prior work in VR network analysis and transitively over 2D screen-based network analysis tools. We attribute these results to the fact that prior VR work has not made extensive use of the available space and that there is great potential in both spatial arrangement, choice of layouts, and interaction design. We hope to provide a direction for future research in this area since there are still several unexplored areas, as hinted at in Table~\ref{tab:task_layout_matrix}.
\section{Limitations \& Future Work}

We believe that the multi-layout method presented here is only the first step toward understanding the potential of interactive, immersive graph layouts. We presented a combination of four possible views, which we show to perform quite well. Although we make informed choices in our design, we do not claim this particular arrangement of views to be flawless. The presented approach was shown to outperform prior work in immersive network visualization, but future studies need to further investigate the effect of various layout methods and different combinations thereof (Table~\ref{tab:task_layout_matrix}).

One limitation of the current evaluation was the fixed FOV and static placement of the layouts. This results in some overwhelmingly big presentations of smaller networks while it creates clutter in more dense datasets. It also leads users to assume uncomfortable head positions when performing certain tasks. These restrictions increase the risk of fatigue which could impair performance and enjoyment of using the system. This issue is reflected in user feedback, as it is often one of the first concerns mentioned by our participants. Future work can address this drawback by adjusting the FOV to achieve homogeneous node and community density and by allowing the user to adjust layout positions manually.

Another possible direction this work can be expanded on is by introducing more intricate interaction methods like gestures~\cite{Erra2019,Huang2017} or eye-tracking to accommodate certain tasks and reduce fagitue~\cite{Masopust2021}.
\section{Conclusion}
We introduce a multi-layout approach to networks in VR. Through this work, we hope to enrich existing efforts to prove the potential of immersive visualization for network tasks. The literature on interactions, layout, and rendering of networks in VR environments is ever-growing. We contribute to this body of work by introducing a novel and effective design for network analysis in immersive space which covers a broad range of layouts and supports a vast array of common network tasks.

Our study results show that the multi-layout approach outperforms the state-of-the-art of immersive network visualization---especially for complex and spatially demanding tasks in hierarchical network structures. Building on top of users' feedback about interaction and layout methods, the next step is to refine our system for further studies. Specifically, we want to further evaluate scenarios that were not covered by our current design (Table~\ref{tab:task_layout_matrix} yellow cells) and extend its capabilities to investigate layout-task combinations that were previously not considered by literature (Table~\ref{tab:task_layout_matrix} red cells). Additionally, more studies on the efficacy of different views, interactions, and layout suite sizes will become necessary as we explore the potential uses of immersive visualizations further. Understanding the suitability of different representations and interactions for this type of data is crucial to driving the design research of information visualization in virtual reality.

% \begin{acks}
% \end{acks}

\bibliographystyle{ACM-Reference-Format}
\bibliography{sample-base}


\begin{thebibliography}{66}

%%% the Web address.

\ifx \showCODEN    \undefined \def \showCODEN     #1{\unskip}     \fi
\ifx \showDOI      \undefined \def \showDOI       #1{#1}\fi
\ifx \showISBNx    \undefined \def \showISBNx     #1{\unskip}     \fi
\ifx \showISBNxiii \undefined \def \showISBNxiii  #1{\unskip}     \fi
\ifx \showISSN     \undefined \def \showISSN      #1{\unskip}     \fi
\ifx \showLCCN     \undefined \def \showLCCN      #1{\unskip}     \fi
\ifx \shownote     \undefined \def \shownote      #1{#1}          \fi
\ifx \showarticletitle \undefined \def \showarticletitle #1{#1}   \fi
\ifx \showURL      \undefined \def \showURL       {\relax}        \fi
% invisible to TeX
\providecommand\bibfield[2]{#2}
\providecommand\bibinfo[2]{#2}
\providecommand\natexlab[1]{#1}
\providecommand\showeprint[2][]{arXiv:#2}

\bibitem[Alper et~al\mbox{.}(2011)]%
        {Alper2011}
\bibfield{author}{\bibinfo{person}{Basak Alper}, \bibinfo{person}{Tobias
  H{\"{o}}llerer}, \bibinfo{person}{Joann Kuchera-Morin}, {and}
  \bibinfo{person}{Angus Forbes}.} \bibinfo{year}{2011}\natexlab{}.
\newblock \showarticletitle{{Stereoscopic highlighting: 2D graph visualization
  on stereo displays}}.
\newblock \bibinfo{journal}{\emph{IEEE Transactions on Visualization and
  Computer Graphics}} \bibinfo{volume}{17}, \bibinfo{number}{12}
  (\bibinfo{year}{2011}), \bibinfo{pages}{2325--2333}.
\newblock
\showISSN{10772626}
\urldef\tempurl%
\url{https://doi.org/10.1109/TVCG.2011.234}
\showDOI{\tempurl}


\bibitem[Amar et~al\mbox{.}(2005)]%
        {amar2005low}
\bibfield{author}{\bibinfo{person}{Robert Amar}, \bibinfo{person}{James Eagan},
  {and} \bibinfo{person}{John Stasko}.} \bibinfo{year}{2005}\natexlab{}.
\newblock \showarticletitle{Low-level components of analytic activity in
  information visualization}. In \bibinfo{booktitle}{\emph{IEEE Symposium on
  Information Visualization, 2005. INFOVIS 2005.}} IEEE,
  \bibinfo{pages}{111--117}.
\newblock
\urldef\tempurl%
\url{https://doi.org/10.1109/INFVIS.2005.1532136}
\showDOI{\tempurl}


\bibitem[Barahimi and Wismath(2014)]%
        {Barahimi2014}
\bibfield{author}{\bibinfo{person}{Farshad Barahimi} {and}
  \bibinfo{person}{Stephen Wismath}.} \bibinfo{year}{2014}\natexlab{}.
\newblock \showarticletitle{{3D graph visualization with the Oculus Rift}}.
\newblock \bibinfo{journal}{\emph{Lecture Notes in Computer Science (including
  subseries Lecture Notes in Artificial Intelligence and Lecture Notes in
  Bioinformatics)}} \bibinfo{volume}{8871}, \bibinfo{number}{3}
  (\bibinfo{year}{2014}), \bibinfo{pages}{519--520}.
\newblock
\showISSN{16113349}


\bibitem[Bennett et~al\mbox{.}(2015)]%
        {Bennett2015}
\bibfield{author}{\bibinfo{person}{Rebecca Bennett}, \bibinfo{person}{David~J.
  Zielinski}, {and} \bibinfo{person}{Regis Kopper}.}
  \bibinfo{year}{2015}\natexlab{}.
\newblock \showarticletitle{{Comparison of interactive environments for the
  archaeological exploration of 3D landscape data}}.
\newblock \bibinfo{journal}{\emph{2014 IEEE VIS International Workshop on
  3DVis, 3DVis 2014}} (\bibinfo{year}{2015}), \bibinfo{pages}{67--71}.
\newblock
\showISBNx{9781479968268}
\urldef\tempurl%
\url{https://doi.org/10.1109/3DVis.2014.7160103}
\showDOI{\tempurl}


\bibitem[Blondel et~al\mbox{.}(2008)]%
        {blondel2008fast}
\bibfield{author}{\bibinfo{person}{Vincent~D Blondel},
  \bibinfo{person}{Jean-Loup Guillaume}, \bibinfo{person}{Renaud Lambiotte},
  {and} \bibinfo{person}{Etienne Lefebvre}.} \bibinfo{year}{2008}\natexlab{}.
\newblock \showarticletitle{Fast unfolding of communities in large networks}.
\newblock \bibinfo{journal}{\emph{Journal of statistical mechanics: theory and
  experiment}} \bibinfo{volume}{2008}, \bibinfo{number}{10}
  (\bibinfo{year}{2008}), \bibinfo{pages}{P10008}.
\newblock
\urldef\tempurl%
\url{https://doi.org/10.1088/1742-5468/2008/10/p10008}
\showDOI{\tempurl}


\bibitem[Bourqui et~al\mbox{.}(2007)]%
        {Bourqui2007}
\bibfield{author}{\bibinfo{person}{Romain Bourqui}, \bibinfo{person}{David
  Auber}, {and} \bibinfo{person}{Patrick Mary}.}
  \bibinfo{year}{2007}\natexlab{}.
\newblock \showarticletitle{{How to draw clustered weighted graphs using a
  multilevel force-directed graph drawing algorithm}}.
\newblock \bibinfo{journal}{\emph{Proceedings of the International Conference
  on Information Visualisation}} (\bibinfo{year}{2007}),
  \bibinfo{pages}{757--764}.
\newblock
\showISSN{10939547}
\urldef\tempurl%
\url{https://doi.org/10.1109/IV.2007.65}
\showDOI{\tempurl}


\bibitem[Bowman et~al\mbox{.}(2004)]%
        {bowman20043d}
\bibfield{author}{\bibinfo{person}{Doug Bowman}, \bibinfo{person}{Ernst
  Kruijff}, \bibinfo{person}{Joseph~J LaViola~Jr}, {and}
  \bibinfo{person}{Ivan~P Poupyrev}.} \bibinfo{year}{2004}\natexlab{}.
\newblock \bibinfo{booktitle}{\emph{3D User interfaces: theory and practice,
  CourseSmart eTextbook}}.
\newblock \bibinfo{publisher}{Addison-Wesley}.
\newblock


\bibitem[Brooks(1999)]%
        {Brooks1999}
\bibfield{author}{\bibinfo{person}{Frederick~P. Brooks}.}
  \bibinfo{year}{1999}\natexlab{}.
\newblock \showarticletitle{{What's real about virtual reality?}}
\newblock \bibinfo{journal}{\emph{IEEE Computer Graphics and Applications}}
  \bibinfo{volume}{19}, \bibinfo{number}{6} (\bibinfo{year}{1999}),
  \bibinfo{pages}{16--27}.
\newblock
\showISSN{02721716}
\urldef\tempurl%
\url{https://doi.org/10.1109/38.799723}
\showDOI{\tempurl}


\bibitem[Bryson(1993)]%
        {Bryson1993}
\bibfield{author}{\bibinfo{person}{Steve Bryson}.}
  \bibinfo{year}{1993}\natexlab{}.
\newblock \showarticletitle{{Virtual reality in scientific visualization}}.
\newblock \bibinfo{journal}{\emph{Computers and Graphics}}
  \bibinfo{volume}{17}, \bibinfo{number}{6} (\bibinfo{year}{1993}),
  \bibinfo{pages}{679--685}.
\newblock
\showISSN{00978493}
\urldef\tempurl%
\url{https://doi.org/10.1016/0097-8493(93)90117-R}
\showDOI{\tempurl}


\bibitem[Buschel et~al\mbox{.}(2019)]%
        {Bueschel2019}
\bibfield{author}{\bibinfo{person}{W. Buschel}, \bibinfo{person}{S. Vogt},
  {and} \bibinfo{person}{R. Dachselt}.} \bibinfo{year}{2019}\natexlab{}.
\newblock \showarticletitle{Augmented Reality Graph Visualizations}.
\newblock \bibinfo{journal}{\emph{IEEE Computer Graphics and Applications}}
  \bibinfo{volume}{39}, \bibinfo{number}{03} (\bibinfo{date}{may}
  \bibinfo{year}{2019}), \bibinfo{pages}{29--40}.
\newblock
\showISSN{1558-1756}
\urldef\tempurl%
\url{https://doi.org/10.1109/MCG.2019.2897927}
\showDOI{\tempurl}


\bibitem[{Cordeil} et~al\mbox{.}(2017)]%
        {Cordeil2017}
\bibfield{author}{\bibinfo{person}{M. {Cordeil}}, \bibinfo{person}{T. {Dwyer}},
  \bibinfo{person}{K. {Klein}}, \bibinfo{person}{B. {Laha}},
  \bibinfo{person}{K. {Marriott}}, {and} \bibinfo{person}{B.~H. {Thomas}}.}
  \bibinfo{year}{2017}\natexlab{}.
\newblock \showarticletitle{Immersive Collaborative Analysis of Network
  Connectivity: CAVE-style or Head-Mounted Display?}
\newblock \bibinfo{journal}{\emph{IEEE Transactions on Visualization and
  Computer Graphics}} \bibinfo{volume}{23}, \bibinfo{number}{1}
  (\bibinfo{year}{2017}), \bibinfo{pages}{441--450}.
\newblock
\urldef\tempurl%
\url{https://doi.org/10.1109/TVCG.2016.2599107}
\showDOI{\tempurl}


\bibitem[Cumming(2014)]%
        {AntiNHST2}
\bibfield{author}{\bibinfo{person}{Geoff Cumming}.}
  \bibinfo{year}{2014}\natexlab{}.
\newblock \showarticletitle{The New Statistics: Why and How}.
\newblock \bibinfo{journal}{\emph{Psychological Science}} \bibinfo{volume}{25},
  \bibinfo{number}{1} (\bibinfo{year}{2014}), \bibinfo{pages}{7--29}.
\newblock
\urldef\tempurl%
\url{https://doi.org/10.1177/0956797613504966}
\showDOI{\tempurl}


\bibitem[Cutting and Vishton(1995)]%
        {cutting1995perceiving}
\bibfield{author}{\bibinfo{person}{James~E Cutting} {and}
  \bibinfo{person}{Peter~M Vishton}.} \bibinfo{year}{1995}\natexlab{}.
\newblock \showarticletitle{Perceiving layout and knowing distances: The
  integration, relative potency, and contextual use of different information
  about depth}.
\newblock In \bibinfo{booktitle}{\emph{Perception of space and motion}}.
  \bibinfo{publisher}{Elsevier}, \bibinfo{pages}{69--117}.
\newblock
\urldef\tempurl%
\url{https://doi.org/10.1016/B978-012240530-3/50005-5}
\showDOI{\tempurl}


\bibitem[Didimo and Montecchiani(2014)]%
        {Didimo2014}
\bibfield{author}{\bibinfo{person}{Walter Didimo} {and}
  \bibinfo{person}{Fabrizio Montecchiani}.} \bibinfo{year}{2014}\natexlab{}.
\newblock \showarticletitle{{Fast layout computation of clustered networks:
  Algorithmic advances and experimental analysis}}.
\newblock \bibinfo{journal}{\emph{Information Sciences}}  \bibinfo{volume}{260}
  (\bibinfo{year}{2014}), \bibinfo{pages}{185--199}.
\newblock
\showISSN{00200255}
\urldef\tempurl%
\url{https://doi.org/10.1016/j.ins.2013.09.048}
\showDOI{\tempurl}


\bibitem[Drogemuller et~al\mbox{.}(2019)]%
        {Drogemueller2019}
\bibfield{author}{\bibinfo{person}{Adam Drogemuller}, \bibinfo{person}{Andrew
  Cunningham}, \bibinfo{person}{James Walsh}, \bibinfo{person}{Bruce Thomas},
  \bibinfo{person}{Maxime Cordeil}, {and} \bibinfo{person}{William Ross}.}
  \bibinfo{year}{2019}\natexlab{}.
\newblock \showarticletitle{Examining Virtual Reality Navigation Techniques for
  3D Network Visualisations}.
\newblock \bibinfo{journal}{\emph{Journal of Visual Languages \& Computing}}
  \bibinfo{volume}{56} (\bibinfo{date}{12} \bibinfo{year}{2019}).
\newblock
\urldef\tempurl%
\url{https://doi.org/10.1016/j.cola.2019.100937}
\showDOI{\tempurl}


\bibitem[Erra et~al\mbox{.}(2019)]%
        {Erra2019}
\bibfield{author}{\bibinfo{person}{Ugo Erra}, \bibinfo{person}{Delfina
  Malandrino}, {and} \bibinfo{person}{Luca Pepe}.}
  \bibinfo{year}{2019}\natexlab{}.
\newblock \showarticletitle{{Virtual Reality Interfaces for Interacting with
  Three-Dimensional Graphs}}.
\newblock \bibinfo{journal}{\emph{International Journal of Human-Computer
  Interaction}} \bibinfo{volume}{35}, \bibinfo{number}{1}
  (\bibinfo{year}{2019}), \bibinfo{pages}{75--88}.
\newblock
\showISSN{15327590}
\urldef\tempurl%
\url{https://doi.org/10.1080/10447318.2018.1429061}
\showDOI{\tempurl}


\bibitem[Ersoy et~al\mbox{.}(2011)]%
        {Ersoy2011}
\bibfield{author}{\bibinfo{person}{Ozan Ersoy}, \bibinfo{person}{Christophe
  Hurter}, \bibinfo{person}{Fernando Paulovich}, \bibinfo{person}{Gabriel
  Cantareiro}, {and} \bibinfo{person}{Alex Telea}.}
  \bibinfo{year}{2011}\natexlab{}.
\newblock \showarticletitle{Skeleton-Based Edge Bundling for Graph
  Visualization}.
\newblock \bibinfo{journal}{\emph{IEEE Transactions on Visualization and
  Computer Graphics}} \bibinfo{volume}{17}, \bibinfo{number}{12}
  (\bibinfo{year}{2011}), \bibinfo{pages}{2364--2373}.
\newblock
\urldef\tempurl%
\url{https://doi.org/10.1109/TVCG.2011.233}
\showDOI{\tempurl}


\bibitem[{Facebook Technologies, LLC.}(2020)]%
        {OculusVR}
\bibfield{author}{\bibinfo{person}{{Facebook Technologies, LLC.}}}
  \bibinfo{year}{2020}\natexlab{}.
\newblock \bibinfo{title}{Oculus | Headsets \& Equipment}.
\newblock \bibinfo{howpublished}{\url{https://www.oculus.com/}}.
\newblock
\newblock
\shownote{[Online; accessed 22-June-2020]}.


\bibitem[Fortunato(2010)]%
        {fortunato2010community}
\bibfield{author}{\bibinfo{person}{Santo Fortunato}.}
  \bibinfo{year}{2010}\natexlab{}.
\newblock \showarticletitle{Community detection in graphs}.
\newblock \bibinfo{journal}{\emph{Physics Reports}} \bibinfo{volume}{486},
  \bibinfo{number}{3-5} (\bibinfo{year}{2010}), \bibinfo{pages}{75--174}.
\newblock
\urldef\tempurl%
\url{https://doi.org/10.1016/j.physrep.2009.11.002}
\showDOI{\tempurl}


\bibitem[Greffard et~al\mbox{.}(2015)]%
        {Greffard2015}
\bibfield{author}{\bibinfo{person}{Nicolas Greffard}, \bibinfo{person}{Fabien
  Picarougne}, {and} \bibinfo{person}{Pascale Kuntz}.}
  \bibinfo{year}{2015}\natexlab{}.
\newblock \showarticletitle{{Beyond the classical monoscopic 3D in graph
  analytics: An experimental study of the impact of stereoscopy}}.
\newblock \bibinfo{journal}{\emph{2014 IEEE VIS International Workshop on
  3DVis, 3DVis 2014}} (\bibinfo{year}{2015}), \bibinfo{pages}{19--24}.
\newblock
\showISBNx{9781479968268}
\urldef\tempurl%
\url{https://doi.org/10.1109/3DVis.2014.7160095}
\showDOI{\tempurl}


\bibitem[Halpin et~al\mbox{.}(2008)]%
        {Halpin2008}
\bibfield{author}{\bibinfo{person}{Harry Halpin}, \bibinfo{person}{David~J.
  Zielinski}, \bibinfo{person}{Rachael Brady}, {and} \bibinfo{person}{Glenda
  Kelly}.} \bibinfo{year}{2008}\natexlab{}.
\newblock \showarticletitle{{Exploring semantic social networks using virtual
  reality}}.
\newblock \bibinfo{journal}{\emph{Lecture Notes in Computer Science (including
  subseries Lecture Notes in Artificial Intelligence and Lecture Notes in
  Bioinformatics)}}  \bibinfo{volume}{5318 LNCS} (\bibinfo{year}{2008}),
  \bibinfo{pages}{599--614}.
\newblock
\showISBNx{3540885633}
\showISSN{03029743}
\urldef\tempurl%
\url{https://doi.org/10.1007/978-3-540-88564-1-38}
\showDOI{\tempurl}


\bibitem[Hanel et~al\mbox{.}(2015)]%
        {Hanel2015}
\bibfield{author}{\bibinfo{person}{Claudia Hanel}, \bibinfo{person}{Benjamin
  Weyers}, \bibinfo{person}{Bernd Hentschel}, {and} \bibinfo{person}{Torsten~W.
  Kuhlen}.} \bibinfo{year}{2015}\natexlab{}.
\newblock \showarticletitle{{Interactive volume rendering for immersive virtual
  environments}}.
\newblock \bibinfo{journal}{\emph{2014 IEEE VIS International Workshop on
  3DVis, 3DVis 2014}}  \bibinfo{volume}{73} (\bibinfo{year}{2015}),
  \bibinfo{pages}{73--74}.
\newblock
\showISBNx{9781479968268}
\urldef\tempurl%
\url{https://doi.org/10.1109/3DVis.2014.7160104}
\showDOI{\tempurl}


\bibitem[Harlow et~al\mbox{.}(2016)]%
        {AntiNHST}
\bibfield{author}{\bibinfo{person}{L. Harlow}, \bibinfo{person}{S. Mulaik},
  {and} \bibinfo{person}{J. Steiger}.} \bibinfo{year}{2016}\natexlab{}.
\newblock \bibinfo{booktitle}{\emph{What If There Were No Significance Tests?}}
\newblock \bibinfo{publisher}{Routledge}, Chapter Eight Common but False
  Objections to the Discontinuation of Significance Testing in the Analysis of
  Research Data.
\newblock


\bibitem[Hart and Staveland(1988)]%
        {hart1988NASATLX}
\bibfield{author}{\bibinfo{person}{Sandra~G Hart} {and}
  \bibinfo{person}{Lowell~E Staveland}.} \bibinfo{year}{1988}\natexlab{}.
\newblock \showarticletitle{Development of NASA-TLX (Task Load Index): Results
  of empirical and theoretical research}.
\newblock In \bibinfo{booktitle}{\emph{Advances in psychology}}.
  Vol.~\bibinfo{volume}{52}. \bibinfo{publisher}{Elsevier},
  \bibinfo{pages}{139--183}.
\newblock
\urldef\tempurl%
\url{https://doi.org/10.1016/S0166-4115(08)62386-9}
\showDOI{\tempurl}


\bibitem[Herman et~al\mbox{.}(2000)]%
        {herman2000graph}
\bibfield{author}{\bibinfo{person}{Ivan Herman}, \bibinfo{person}{Guy
  Melan{\c{c}}on}, {and} \bibinfo{person}{M~Scott Marshall}.}
  \bibinfo{year}{2000}\natexlab{}.
\newblock \showarticletitle{Graph visualization and navigation in information
  visualization: A survey}.
\newblock \bibinfo{journal}{\emph{IEEE Transactions on Visualization and
  Computer Graphics}} \bibinfo{volume}{6}, \bibinfo{number}{1}
  (\bibinfo{year}{2000}), \bibinfo{pages}{24--43}.
\newblock
\urldef\tempurl%
\url{https://doi.org/10.1109/2945.841119}
\showDOI{\tempurl}


\bibitem[Holten(2006)]%
        {Holten2006}
\bibfield{author}{\bibinfo{person}{Danny Holten}.}
  \bibinfo{year}{2006}\natexlab{}.
\newblock \showarticletitle{{Hierarchical edge bundles: Visualization of
  adjacency relations in hierarchical data}}.
\newblock \bibinfo{journal}{\emph{IEEE Transactions on Visualization and
  Computer Graphics}} \bibinfo{volume}{12}, \bibinfo{number}{5}
  (\bibinfo{year}{2006}), \bibinfo{pages}{741--748}.
\newblock
\showISSN{10772626}
\urldef\tempurl%
\url{https://doi.org/10.1109/TVCG.2006.147}
\showDOI{\tempurl}


\bibitem[Holten and van Wijk(2009)]%
        {Holten2009}
\bibfield{author}{\bibinfo{person}{Danny Holten} {and} \bibinfo{person}{Jarke
  van Wijk}.} \bibinfo{year}{2009}\natexlab{}.
\newblock \showarticletitle{Force-Directed Edge Bundling for Graph
  Visualization}.
\newblock \bibinfo{journal}{\emph{Comput. Graph. Forum}}  \bibinfo{volume}{28}
  (\bibinfo{date}{06} \bibinfo{year}{2009}), \bibinfo{pages}{983--990}.
\newblock
\urldef\tempurl%
\url{https://doi.org/10.1111/j.1467-8659.2009.01450.x}
\showDOI{\tempurl}


\bibitem[{HTC Corporation}(2020)]%
        {VIVETM}
\bibfield{author}{\bibinfo{person}{{HTC Corporation}}.}
  \bibinfo{year}{2020}\natexlab{}.
\newblock \bibinfo{title}{VIVE}.
\newblock \bibinfo{howpublished}{\url{https://www.vive.com/us/}}.
\newblock
\newblock
\shownote{[Online; accessed 22-June-2020]}.


\bibitem[Huang et~al\mbox{.}(2017)]%
        {Huang2017}
\bibfield{author}{\bibinfo{person}{Yi~Jheng Huang}, \bibinfo{person}{Takanori
  Fujiwara}, \bibinfo{person}{Yun~Xuan Lin}, \bibinfo{person}{Wen~Chieh Lin},
  {and} \bibinfo{person}{Kwan~Liu Ma}.} \bibinfo{year}{2017}\natexlab{}.
\newblock \showarticletitle{{A gesture system for graph visualization in
  virtual reality environments}}.
\newblock \bibinfo{journal}{\emph{IEEE Pacific Visualization Symposium}}
  (\bibinfo{year}{2017}), \bibinfo{pages}{41--45}.
\newblock
\showISBNx{9781509057382}
\showISSN{21658773}
\urldef\tempurl%
\url{https://doi.org/10.1109/PACIFICVIS.2017.8031577}
\showDOI{\tempurl}


\bibitem[{Hurter} et~al\mbox{.}(2019)]%
        {Hurter2019}
\bibfield{author}{\bibinfo{person}{C. {Hurter}}, \bibinfo{person}{N.~H.
  {Riche}}, \bibinfo{person}{S.~M. {Drucker}}, \bibinfo{person}{M. {Cordeil}},
  \bibinfo{person}{R. {Alligier}}, {and} \bibinfo{person}{R. {Vuillemot}}.}
  \bibinfo{year}{2019}\natexlab{}.
\newblock \showarticletitle{FiberClay: Sculpting Three Dimensional Trajectories
  to Reveal Structural Insights}.
\newblock \bibinfo{journal}{\emph{IEEE Transactions on Visualization and
  Computer Graphics}} \bibinfo{volume}{25}, \bibinfo{number}{1}
  (\bibinfo{year}{2019}), \bibinfo{pages}{704--714}.
\newblock
\urldef\tempurl%
\url{https://doi.org/10.1109/TVCG.2018.2865191}
\showDOI{\tempurl}


\bibitem[Itoh et~al\mbox{.}(2009)]%
        {Itoh2009}
\bibfield{author}{\bibinfo{person}{Takayuki Itoh}, \bibinfo{person}{Chris
  Muelder}, \bibinfo{person}{Kwan-Liu Ma}, {and} \bibinfo{person}{Jun Sese}.}
  \bibinfo{year}{2009}\natexlab{}.
\newblock \showarticletitle{A hybrid space-filling and force-directed layout
  method for visualizing multiple-category graphs}. In
  \bibinfo{booktitle}{\emph{2009 IEEE Pacific Visualization Symposium}}. IEEE,
  \bibinfo{pages}{121--128}.
\newblock
\showISSN{2165-8773}
\urldef\tempurl%
\url{https://doi.org/10.1109/PACIFICVIS.2009.4906846}
\showDOI{\tempurl}


\bibitem[Jacomy et~al\mbox{.}(2014)]%
        {forceatlas2}
\bibfield{author}{\bibinfo{person}{Mathieu Jacomy}, \bibinfo{person}{Tommaso
  Venturini}, \bibinfo{person}{Sebastien Heymann}, {and}
  \bibinfo{person}{Mathieu Bastian}.} \bibinfo{year}{2014}\natexlab{}.
\newblock \showarticletitle{ForceAtlas2, a Continuous Graph Layout Algorithm
  for Handy Network Visualization Designed for the Gephi Software}.
\newblock \bibinfo{journal}{\emph{PLOS ONE}} \bibinfo{volume}{9},
  \bibinfo{number}{6} (\bibinfo{date}{06} \bibinfo{year}{2014}),
  \bibinfo{pages}{1--12}.
\newblock
\urldef\tempurl%
\url{https://doi.org/10.1371/journal.pone.0098679}
\showDOI{\tempurl}


\bibitem[Johnson and Shneiderman(1991)]%
        {Johnson1991}
\bibfield{author}{\bibinfo{person}{Brian Johnson} {and} \bibinfo{person}{Ben
  Shneiderman}.} \bibinfo{year}{1991}\natexlab{}.
\newblock \showarticletitle{{Tree-maps: A space-filling approach to the
  visualization of hierarchical information structures}}.
\newblock \bibinfo{journal}{\emph{Proceedings of the 2nd Conference on
  Visualization 1991, VIS 1991}} (\bibinfo{year}{1991}),
  \bibinfo{pages}{284--291}.
\newblock
\showISBNx{0818622458}
\urldef\tempurl%
\url{https://doi.org/10.1109/visual.1991.175815}
\showDOI{\tempurl}


\bibitem[Kobourov(2012)]%
        {Kobourov2012}
\bibfield{author}{\bibinfo{person}{Stephen~G. Kobourov}.}
  \bibinfo{year}{2012}\natexlab{}.
\newblock \showarticletitle{{Spring Embedders and Force Directed Graph Drawing
  Algorithms}}.
\newblock  (\bibinfo{year}{2012}), \bibinfo{pages}{1--23}.
\newblock
\showeprint[arxiv]{1201.3011}
\urldef\tempurl%
\url{http://arxiv.org/abs/1201.3011}
\showURL{%
\tempurl}


\bibitem[Kotlarek et~al\mbox{.}(2020)]%
        {Kotlarek2020}
\bibfield{author}{\bibinfo{person}{Joseph Kotlarek}, \bibinfo{person}{Oh-Hyun
  Kwon}, \bibinfo{person}{Kwan~Liu Ma}, \bibinfo{person}{Peter Eades},
  \bibinfo{person}{Andreas Kerren}, \bibinfo{person}{Karsten Klein}, {and}
  \bibinfo{person}{Falk Schreiber}.} \bibinfo{year}{2020}\natexlab{}.
\newblock \showarticletitle{{A Study of Mental Maps in Immersive Network
  Visualization}}.
\newblock \bibinfo{journal}{\emph{IEEE Pacific Visualization Symposium}}
  \bibinfo{volume}{2020-June} (\bibinfo{year}{2020}), \bibinfo{pages}{1--10}.
\newblock
\showISBNx{9781728156972}
\showISSN{21658773}
\urldef\tempurl%
\url{https://doi.org/10.1109/PacificVis48177.2020.4722}
\showDOI{\tempurl}
\showeprint[arxiv]{2001.06462}


\bibitem[Kwon et~al\mbox{.}(2016)]%
        {Kwon2016}
\bibfield{author}{\bibinfo{person}{Oh-Hyun Kwon}, \bibinfo{person}{Chris
  Muelder}, \bibinfo{person}{Kyungwon Lee}, {and} \bibinfo{person}{Kwan~Liu
  Ma}.} \bibinfo{year}{2016}\natexlab{}.
\newblock \showarticletitle{{A study of layout, rendering, and interaction
  methods for immersive graph visualization}}.
\newblock \bibinfo{journal}{\emph{IEEE Transactions on Visualization and
  Computer Graphics}} \bibinfo{volume}{22}, \bibinfo{number}{7}
  (\bibinfo{year}{2016}), \bibinfo{pages}{1802--1815}.
\newblock
\showISSN{10772626}
\urldef\tempurl%
\url{https://doi.org/10.1109/TVCG.2016.2520921}
\showDOI{\tempurl}


\bibitem[Lee et~al\mbox{.}(2006)]%
        {Lee2006}
\bibfield{author}{\bibinfo{person}{Bongshin Lee}, \bibinfo{person}{Catherine
  Plaisant}, \bibinfo{person}{Jean-Daniel Fekete},
  \bibinfo{person}{Cynthia~Sims Parr}, {and} \bibinfo{person}{Nathalie Henry}.}
  \bibinfo{year}{2006}\natexlab{}.
\newblock \showarticletitle{{Task Taxonomy for Graph Visualization Categories
  and Subject Descriptors}}.
\newblock \bibinfo{journal}{\emph{Proceedings of the AVI Workshop Beyond Time
  and Errors: Novel Evaluation Methods for Information Visualization}}
  (\bibinfo{year}{2006}), \bibinfo{pages}{1--5}.
\newblock
\showISBNx{1301405744}
\urldef\tempurl%
\url{https://doi.org/10.1145/1168149.1168168}
\showDOI{\tempurl}


\bibitem[Marriott et~al\mbox{.}(2018)]%
        {Marriott2018}
\bibfield{author}{\bibinfo{person}{Kim Marriott}, \bibinfo{person}{Jian Chen},
  \bibinfo{person}{Marcel Hlawatsch}, \bibinfo{person}{Takayuki Itoh},
  \bibinfo{person}{Miguel~A. Nacenta}, \bibinfo{person}{Guido Reina}, {and}
  \bibinfo{person}{Wolfgang Stuerzlinger}.} \bibinfo{year}{2018}\natexlab{}.
\newblock \bibinfo{booktitle}{\emph{Immersive Analytics: Time to Reconsider the
  Value of 3D for Information Visualisation}}.
\newblock \bibinfo{publisher}{Springer International Publishing},
  \bibinfo{address}{Cham}, \bibinfo{pages}{25--55}.
\newblock
\showISBNx{978-3-030-01388-2}
\urldef\tempurl%
\url{https://doi.org/10.1007/978-3-030-01388-2}
\showDOI{\tempurl}


\bibitem[Masopust et~al\mbox{.}(2021)]%
        {Masopust2021}
\bibfield{author}{\bibinfo{person}{Lukas~Maximilian Masopust},
  \bibinfo{person}{David Bauer}, \bibinfo{person}{Siyuan Yao}, {and}
  \bibinfo{person}{Kwan-Liu Ma}.} \bibinfo{year}{2021}\natexlab{}.
\newblock \showarticletitle{A Comparison of the Fatigue Progression of
  Eye-Tracked and Motion-Controlled Interaction in Immersive Space}. In
  \bibinfo{booktitle}{\emph{2021 IEEE International Symposium on Mixed and
  Augmented Reality (ISMAR)}}. \bibinfo{pages}{460--469}.
\newblock
\urldef\tempurl%
\url{https://doi.org/10.1109/ISMAR52148.2021.00063}
\showDOI{\tempurl}


\bibitem[McIntire and Liggett(2015)]%
        {McIntire2015}
\bibfield{author}{\bibinfo{person}{John~P. McIntire} {and}
  \bibinfo{person}{Kristen~K. Liggett}.} \bibinfo{year}{2015}\natexlab{}.
\newblock \showarticletitle{{The (possible) utility of stereoscopic 3D displays
  for information visualization: The good, the bad, and the ugly}}.
\newblock \bibinfo{journal}{\emph{2014 IEEE VIS International Workshop on
  3DVis, 3DVis 2014}} (\bibinfo{year}{2015}), \bibinfo{pages}{1--9}.
\newblock
\showISBNx{9781479968268}
\urldef\tempurl%
\url{https://doi.org/10.1109/3DVis.2014.7160093}
\showDOI{\tempurl}


\bibitem[Miller(1956)]%
        {MillerMagic1956}
\bibfield{author}{\bibinfo{person}{George~A. Miller}.}
  \bibinfo{year}{1956}\natexlab{}.
\newblock \showarticletitle{The magical number seven, plus or minus two: some
  limits on our capacity for processing information.}
\newblock \bibinfo{journal}{\emph{Psychological Review}} \bibinfo{volume}{63},
  \bibinfo{number}{2} (\bibinfo{year}{1956}), \bibinfo{pages}{81--97}.
\newblock
\urldef\tempurl%
\url{https://doi.org/10.1037/h0043158}
\showDOI{\tempurl}


\bibitem[Mirhosseini et~al\mbox{.}(2015)]%
        {Mirhosseini2015}
\bibfield{author}{\bibinfo{person}{Koosha Mirhosseini}, \bibinfo{person}{Qi
  Sun}, \bibinfo{person}{Krishna~C. Gurijala}, \bibinfo{person}{Bireswar Laha},
  {and} \bibinfo{person}{Arie~E. Kaufman}.} \bibinfo{year}{2015}\natexlab{}.
\newblock \showarticletitle{{Benefits of 3D immersion for virtual
  colonoscopy}}.
\newblock \bibinfo{journal}{\emph{2014 IEEE VIS International Workshop on
  3DVis, 3DVis 2014}} \bibinfo{number}{Vc} (\bibinfo{year}{2015}),
  \bibinfo{pages}{75--79}.
\newblock
\showISBNx{9781479968268}
\urldef\tempurl%
\url{https://doi.org/10.1109/3DVis.2014.7160105}
\showDOI{\tempurl}


\bibitem[Mirhosseini et~al\mbox{.}(2019)]%
        {Mirhosseini2019}
\bibfield{author}{\bibinfo{person}{Seyedkoosha Mirhosseini},
  \bibinfo{person}{Ievgeniia Gutenko}, \bibinfo{person}{Sushant Ojal},
  \bibinfo{person}{Joseph Marino}, {and} \bibinfo{person}{Arie Kaufman}.}
  \bibinfo{year}{2019}\natexlab{}.
\newblock \showarticletitle{Immersive Virtual Colonoscopy}.
\newblock \bibinfo{journal}{\emph{IEEE Transactions on Visualization and
  Computer Graphics}} \bibinfo{volume}{25}, \bibinfo{number}{5}
  (\bibinfo{year}{2019}), \bibinfo{pages}{2011--2021}.
\newblock
\urldef\tempurl%
\url{https://doi.org/10.1109/TVCG.2019.2898763}
\showDOI{\tempurl}


\bibitem[Munzner(1997)]%
        {Munzner1997}
\bibfield{author}{\bibinfo{person}{Tamara Munzner}.}
  \bibinfo{year}{1997}\natexlab{}.
\newblock \showarticletitle{{H3: laying out large directed graphs in 3D
  hyperbolic space}}.
\newblock \bibinfo{journal}{\emph{Proceedings of the IEEE Symposium on
  Information Visualization}} (\bibinfo{year}{1997}), \bibinfo{pages}{2--10}.
\newblock
\urldef\tempurl%
\url{https://doi.org/10.1109/infvis.1997.636718}
\showDOI{\tempurl}


\bibitem[Phan et~al\mbox{.}(2005)]%
        {Phan2005Flow}
\bibfield{author}{\bibinfo{person}{Doantam Phan}, \bibinfo{person}{Ling Xiao},
  \bibinfo{person}{Ron Yeh}, {and} \bibinfo{person}{Pat Hanrahan}.}
  \bibinfo{year}{2005}\natexlab{}.
\newblock \showarticletitle{Flow map layout}. In \bibinfo{booktitle}{\emph{IEEE
  Symposium on Information Visualization, 2005. INFOVIS 2005.}} IEEE,
  \bibinfo{pages}{219--224}.
\newblock
\urldef\tempurl%
\url{https://doi.org/10.1109/INFVIS.2005.1532150}
\showDOI{\tempurl}


\bibitem[Rekimoto and Green(1993)]%
        {Rekimoto1993}
\bibfield{author}{\bibinfo{person}{Jun Rekimoto} {and} \bibinfo{person}{Mark
  Green}.} \bibinfo{year}{1993}\natexlab{}.
\newblock \showarticletitle{The information cube: Using transparency in 3d
  information visualization}. In \bibinfo{booktitle}{\emph{Proceedings of the
  Third Annual Workshop on Information Technologies \& Systems (WITS’93)}}.
  \bibinfo{pages}{125--132}.
\newblock


\bibitem[Robertson et~al\mbox{.}(1991)]%
        {Robertson1991}
\bibfield{author}{\bibinfo{person}{George~G. Robertson},
  \bibinfo{person}{Jock~D. Mackinlay}, {and} \bibinfo{person}{Stuart~K. Card}.}
  \bibinfo{year}{1991}\natexlab{}.
\newblock \showarticletitle{Cone Trees: Animated 3D Visualizations of
  Hierarchical Information}. In \bibinfo{booktitle}{\emph{Proceedings of the
  SIGCHI Conference on Human Factors in Computing Systems}} (New Orleans,
  Louisiana, USA) \emph{(\bibinfo{series}{CHI ’91})}.
  \bibinfo{publisher}{Association for Computing Machinery},
  \bibinfo{address}{New York, NY, USA}, \bibinfo{pages}{189–194}.
\newblock
\showISBNx{0897913833}
\urldef\tempurl%
\url{https://doi.org/10.1145/108844.108883}
\showDOI{\tempurl}


\bibitem[Saket et~al\mbox{.}(2014)]%
        {Saket2014}
\bibfield{author}{\bibinfo{person}{Bahador Saket}, \bibinfo{person}{Paolo
  Simonetto}, {and} \bibinfo{person}{Stephen Kobourov}.}
  \bibinfo{year}{2014}\natexlab{}.
\newblock \showarticletitle{{Group-Level Graph Visualization Taxonomy}}. In
  \bibinfo{booktitle}{\emph{EuroVis - Short Papers}},
  \bibfield{editor}{\bibinfo{person}{N.~Elmqvist},
  \bibinfo{person}{M.~Hlawitschka}, {and} \bibinfo{person}{J.~Kennedy}} (Eds.).
  \bibinfo{publisher}{The Eurographics Association}.
\newblock
\showISBNx{978-3-905674-69-9}
\urldef\tempurl%
\url{https://doi.org/10.2312/eurovisshort.20141162}
\showDOI{\tempurl}


\bibitem[{SAMSUNG ELECTRONICS CO., LTD.}(2020)]%
        {SamsungGearVR}
\bibfield{author}{\bibinfo{person}{{SAMSUNG ELECTRONICS CO., LTD.}}}
  \bibinfo{year}{2020}\natexlab{}.
\newblock \bibinfo{title}{Samsung Gear}.
\newblock
  \bibinfo{howpublished}{\url{http://www.samsung.com/global/galaxy/gear-vr/}}.
\newblock
\newblock
\shownote{[Online; accessed 22-June-2020]}.


\bibitem[Scholl et~al\mbox{.}(2019)]%
        {Scholl2019}
\bibfield{author}{\bibinfo{person}{Ingrid Scholl}, \bibinfo{person}{Alexander
  Bartella}, \bibinfo{person}{Cem Moluluo}, \bibinfo{person}{Berat Ertural},
  \bibinfo{person}{Frederic Laing}, {and} \bibinfo{person}{Sebastian Suder}.}
  \bibinfo{year}{2019}\natexlab{}.
\newblock \bibinfo{booktitle}{\emph{MedicVR: Acceleration and Enhancement
  Techniques for Direct Volume Rendering in Virtual Reality}}.
\newblock \bibinfo{pages}{152--157}.
\newblock
\urldef\tempurl%
\url{https://doi.org/10.1007/978-3-658-25326-4_32}
\showDOI{\tempurl}


\bibitem[Scholl et~al\mbox{.}(2018)]%
        {Scholl2018}
\bibfield{author}{\bibinfo{person}{Ingrid Scholl}, \bibinfo{person}{Sebastian
  Suder}, {and} \bibinfo{person}{Stefan Schiffer}.}
  \bibinfo{year}{2018}\natexlab{}.
\newblock \showarticletitle{{Direct volume rendering in virtual reality}}.
\newblock \bibinfo{journal}{\emph{Informatik aktuell}} \bibinfo{volume}{d},
  \bibinfo{number}{211279} (\bibinfo{year}{2018}), \bibinfo{pages}{297--302}.
\newblock
\showISSN{1431472X}
\urldef\tempurl%
\url{https://doi.org/10.1007/978-3-662-56537-7_79}
\showDOI{\tempurl}


\bibitem[{Sony Interactive Entertainment LLC.}(2020)]%
        {PlayStationVR}
\bibfield{author}{\bibinfo{person}{{Sony Interactive Entertainment LLC.}}}
  \bibinfo{year}{2020}\natexlab{}.
\newblock \bibinfo{title}{PlayStation}.
\newblock
  \bibinfo{howpublished}{\url{https://www.playstation.com/en-us/explore/playstation-vr/}}.
\newblock
\newblock
\shownote{[Online; accessed 22-June-2020]}.


\bibitem[Sorger et~al\mbox{.}(2019)]%
        {Sorger2019}
\bibfield{author}{\bibinfo{person}{Johannes Sorger}, \bibinfo{person}{Manuela
  Waldner}, \bibinfo{person}{Wolfgang Knecht}, {and} \bibinfo{person}{Alessio
  Arleo}.} \bibinfo{year}{2019}\natexlab{}.
\newblock \showarticletitle{{Immersive analytics of large dynamic networks via
  overview and detail navigation}}.
\newblock \bibinfo{journal}{\emph{Proceedings - 2019 IEEE International
  Conference on Artificial Intelligence and Virtual Reality, AIVR 2019}}
  (\bibinfo{year}{2019}), \bibinfo{pages}{144--151}.
\newblock
\showISBNx{9781728156040}
\urldef\tempurl%
\url{https://doi.org/10.1109/AIVR46125.2019.00030}
\showDOI{\tempurl}
\showeprint[arxiv]{1910.06825}


\bibitem[Sutherland(1965)]%
        {sutherland1965ultimate}
\bibfield{author}{\bibinfo{person}{Ivan~E. Sutherland}.}
  \bibinfo{year}{1965}\natexlab{}.
\newblock \showarticletitle{The Ultimate Display}. In
  \bibinfo{booktitle}{\emph{Proceedings of the IFIP Congress}}.
  \bibinfo{pages}{506--508}.
\newblock


\bibitem[Sweller et~al\mbox{.}(1998)]%
        {Sweller1998}
\bibfield{author}{\bibinfo{person}{John Sweller}, \bibinfo{person}{Jeroen J.~G.
  van Merrienboer}, {and} \bibinfo{person}{Fred G. W.~C. Paas}.}
  \bibinfo{year}{1998}\natexlab{}.
\newblock \showarticletitle{Cognitive Architecture and Instructional Design}.
\newblock \bibinfo{journal}{\emph{Educational Psychology Review}}
  \bibinfo{volume}{10}, \bibinfo{number}{3} (\bibinfo{date}{01 Sep}
  \bibinfo{year}{1998}), \bibinfo{pages}{251--296}.
\newblock
\showISSN{1573-336X}
\urldef\tempurl%
\url{https://doi.org/10.1023/A:1022193728205}
\showDOI{\tempurl}


\bibitem[{Van Dam} et~al\mbox{.}(2000)]%
        {VanDam2000}
\bibfield{author}{\bibinfo{person}{Andries {Van Dam}},
  \bibinfo{person}{Andrew~S. Forsberg}, \bibinfo{person}{David~H. Laidlaw},
  \bibinfo{person}{Joseph~J. LaViola}, {and} \bibinfo{person}{Rosemary~M.
  Simpson}.} \bibinfo{year}{2000}\natexlab{}.
\newblock \showarticletitle{{Immersive VR for scientific visualization: A
  progress report}}.
\newblock \bibinfo{journal}{\emph{IEEE Computer Graphics and Applications}}
  \bibinfo{volume}{20}, \bibinfo{number}{6} (\bibinfo{year}{2000}),
  \bibinfo{pages}{26--52}.
\newblock
\showISSN{02721716}
\urldef\tempurl%
\url{https://doi.org/10.1109/38.888006}
\showDOI{\tempurl}


\bibitem[Vehlow et~al\mbox{.}(2017)]%
        {Vehlow2017}
\bibfield{author}{\bibinfo{person}{Corinna Vehlow}, \bibinfo{person}{Fabian
  Beck}, {and} \bibinfo{person}{Daniel Weiskopf}.}
  \bibinfo{year}{2017}\natexlab{}.
\newblock \showarticletitle{{Visualizing Group Structures in Graphs: A
  Survey}}.
\newblock \bibinfo{journal}{\emph{Computer Graphics Forum}}
  \bibinfo{volume}{36}, \bibinfo{number}{6} (\bibinfo{year}{2017}),
  \bibinfo{pages}{201--225}.
\newblock
\showISSN{14678659}
\urldef\tempurl%
\url{https://doi.org/10.1111/cgf.12872}
\showDOI{\tempurl}


\bibitem[Ware(2019)]%
        {ware2019information}
\bibfield{author}{\bibinfo{person}{Colin Ware}.}
  \bibinfo{year}{2019}\natexlab{}.
\newblock \bibinfo{booktitle}{\emph{Information visualization: perception for
  design}}.
\newblock \bibinfo{publisher}{Morgan Kaufmann}.
\newblock


\bibitem[Ware and Franck(1996)]%
        {Ware1996}
\bibfield{author}{\bibinfo{person}{Colin Ware} {and} \bibinfo{person}{Glenn
  Franck}.} \bibinfo{year}{1996}\natexlab{}.
\newblock \showarticletitle{{Evaluating Stereo and Motion Cues for Visualizing
  Information Nets in Three Dimensions}}.
\newblock \bibinfo{journal}{\emph{ACM Transactions on Graphics}}
  \bibinfo{volume}{15}, \bibinfo{number}{2} (\bibinfo{year}{1996}),
  \bibinfo{pages}{121--140}.
\newblock
\showISSN{07300301}
\urldef\tempurl%
\url{https://doi.org/10.1145/234972.234975}
\showDOI{\tempurl}


\bibitem[Ware and Mitchell(2005)]%
        {Ware2005}
\bibfield{author}{\bibinfo{person}{Colin Ware} {and} \bibinfo{person}{Peter
  Mitchell}.} \bibinfo{year}{2005}\natexlab{}.
\newblock \showarticletitle{Reevaluating Stereo and Motion Cues for Visualizing
  Graphs in Three Dimensions}. In \bibinfo{booktitle}{\emph{Proceedings of the
  2nd Symposium on Applied Perception in Graphics and Visualization}} (A
  Coro\~{n}a, Spain) \emph{(\bibinfo{series}{APGV ’05})}.
  \bibinfo{publisher}{Association for Computing Machinery},
  \bibinfo{address}{New York, NY, USA}, \bibinfo{pages}{51–58}.
\newblock
\showISBNx{1595931392}
\urldef\tempurl%
\url{https://doi.org/10.1145/1080402.1080411}
\showDOI{\tempurl}


\bibitem[Ware and Mitchell(2008)]%
        {Ware2008}
\bibfield{author}{\bibinfo{person}{Colin Ware} {and} \bibinfo{person}{Peter
  Mitchell}.} \bibinfo{year}{2008}\natexlab{}.
\newblock \showarticletitle{Visualizing graphs in three dimensions}.
\newblock \bibinfo{journal}{\emph{ACM Transactions on Applied Perception
  (TAP)}} \bibinfo{volume}{5}, \bibinfo{number}{1} (\bibinfo{year}{2008}),
  \bibinfo{pages}{1--15}.
\newblock


\bibitem[Watts and Strogatz(1998)]%
        {celegansneural}
\bibfield{author}{\bibinfo{person}{Duncan~J. Watts} {and}
  \bibinfo{person}{Steven~H. Strogatz}.} \bibinfo{year}{1998}\natexlab{}.
\newblock \showarticletitle{Collective dynamics of `small-world' networks}.
\newblock \bibinfo{journal}{\emph{Nature}} \bibinfo{volume}{393},
  \bibinfo{number}{6684} (\bibinfo{date}{01 Jun} \bibinfo{year}{1998}),
  \bibinfo{pages}{440--442}.
\newblock
\showISSN{1476-4687}
\urldef\tempurl%
\url{https://doi.org/10.1038/30918}
\showDOI{\tempurl}


\bibitem[Yang et~al\mbox{.}(2013)]%
        {Yang2013}
\bibfield{author}{\bibinfo{person}{Jing Yang}, \bibinfo{person}{Yujie Liu},
  \bibinfo{person}{Xin Zhang}, \bibinfo{person}{Xiaoru Yuan},
  \bibinfo{person}{Ye Zhao}, \bibinfo{person}{Scott Barlowe}, {and}
  \bibinfo{person}{Shixia Liu}.} \bibinfo{year}{2013}\natexlab{}.
\newblock \showarticletitle{{PIWI: Visually exploring graphs based on their
  community structure}}.
\newblock \bibinfo{journal}{\emph{IEEE Transactions on Visualization and
  Computer Graphics}} \bibinfo{volume}{19}, \bibinfo{number}{6}
  (\bibinfo{year}{2013}), \bibinfo{pages}{1034--1047}.
\newblock
\showISSN{10772626}
\urldef\tempurl%
\url{https://doi.org/10.1109/TVCG.2012.172}
\showDOI{\tempurl}


\bibitem[Yoghourdjian et~al\mbox{.}(2018)]%
        {Yoghourdjian2018}
\bibfield{author}{\bibinfo{person}{Vahan Yoghourdjian}, \bibinfo{person}{Daniel
  Archambault}, \bibinfo{person}{Stephan Diehl}, \bibinfo{person}{Tim Dwyer},
  \bibinfo{person}{Karsten Klein}, \bibinfo{person}{Helen~C. Purchase}, {and}
  \bibinfo{person}{Hsiang~Yun Wu}.} \bibinfo{year}{2018}\natexlab{}.
\newblock \showarticletitle{{Exploring the limits of complexity: A survey of
  empirical studies on graph visualisation}}.
\newblock \bibinfo{journal}{\emph{Visual Informatics}} \bibinfo{volume}{2},
  \bibinfo{number}{4} (\bibinfo{year}{2018}), \bibinfo{pages}{264--282}.
\newblock
\showISSN{2468502X}
\urldef\tempurl%
\url{https://doi.org/10.1016/j.visinf.2018.12.006}
\showDOI{\tempurl}
\showeprint[arxiv]{1809.00270}


\bibitem[Youdas et~al\mbox{.}(1992)]%
        {Youdas1992}
\bibfield{author}{\bibinfo{person}{James~W Youdas}, \bibinfo{person}{Tom~R
  Garrett}, \bibinfo{person}{Vera~J Suman}, \bibinfo{person}{Connie~L Bogard},
  \bibinfo{person}{Horace~O Hallman}, {and} \bibinfo{person}{James~R Carey}.}
  \bibinfo{year}{1992}\natexlab{}.
\newblock \showarticletitle{{Normal Range of Motion of the Cervical Spine: An
  Initial Goniometric Study}}.
\newblock \bibinfo{journal}{\emph{Physical Therapy}} \bibinfo{volume}{72},
  \bibinfo{number}{11} (\bibinfo{date}{11} \bibinfo{year}{1992}),
  \bibinfo{pages}{770--780}.
\newblock
\showISSN{0031-9023}
\urldef\tempurl%
\url{https://doi.org/10.1093/ptj/72.11.770}
\showDOI{\tempurl}


\bibitem[Young(1996)]%
        {Young1996}
\bibfield{author}{\bibinfo{person}{Peter Young}.}
  \bibinfo{year}{1996}\natexlab{}.
\newblock \showarticletitle{{Three Dimensional Information Visualization}}.
\newblock \bibinfo{journal}{\emph{Technical Report 12/96, University of
  Durham}} (\bibinfo{year}{1996}).
\newblock


\end{thebibliography}


\end{document}

\title{A Multi-Layout Design for Immersive Visualization of Network Data\\Supplemental Material}

\ccsdesc[500]{Human-centered computing~Graph drawings}
\ccsdesc[300]{Human-centered computing~Information visualization}
\ccsdesc[500]{Human-centered computing~Virtual reality}
\ccsdesc[300]{Human-centered computing~User studies}

\keywords{network community structure, graph layout, head-mounted display, immersive environment}

%% 
%% Teaser Figure
%     \includegraphics[width=\textwidth]{images/header.png}
%     \caption{A view of a hierarchical network rendered with our system. A single group is expanded. Its nodes and intra-group connectivity can be inspected in detail while preserving inter-group context.}
%     \Description{figure description}
%     \label{fig:teaser}

\maketitle

%%
%% Section Includes
% \documentclass[../supplemental.tex]{subfiles}

% \begin{document}
\section{Video}
We provide a supplemental video that describes the system, summarizes the study findings and visually explores the different layouts based on the use-cases described in the main document. Please refer to this video to get a better impression of how our system looks and feels visually.

% \biblio
% \end{document}
%\input{supplemental_sections/usecases}
% \documentclass[../supplemental.tex]{subfiles}

% \begin{document}
\section{Study Data}
Attached to the publication is an archive containing anonymized numerical study data.
Each file corresponds to one participant. All files are provided in CSV format.
The columns are as follows.

\begin{itemize}
    \item \textbf{condition} - The condition used in this trial. Either 'BASELINE' or 'MULTI'.
    \item \textbf{graphID} - The id of the graph used in the task. One of '0', '1', or '2' which corresponds to the graphs \textit{Easy}, \textit{Medium}, and \textit{Hard}.
    \item \textbf{taskID} - the id of the task executed. One of '0', '1', '2', '3', or '4' corresponding to the order of tasks as given in the paper.
    \item \textbf{startTime} - The absolute start time of the task in UNIX time format.
    \item \textbf{endTime} - The absolute end time of the task in UNIX time format.
    \item \textbf{duration} - The relative duration of the task in seconds.
    \item \textbf{correctAnswerProvided} - A boolean indicating if the task was answered 100\% correctly.
    \item \textbf{numberOfInteractions} - Raw, unprocessed number of interactions registered by the system. This is the sum of all types of interactions plus task selection interactions. Due to our system design task interactions are doubly counted. We correct for this factor during data analysis.
    \item \textbf{numberOfExpansions} - Raw, unprocessed number of community expansions registered by the system.
    \item \textbf{numberOfProjections} - Raw, unprocessed number of community projections registered by the system.
    \item \textbf{numberOfOverviews} - Raw, unprocessed number of overview views registered by the system. Note that this value might be non-zero for some entries even in the \textit{BASE} condition as the system automatically switches to this layout between tasks. We correct for this factor during data analysis.
    \item \textbf{numberOfSphericalViews} - Raw, unprocessed number of spherical layout views registered by the system.
    \item \textbf{accuracy} - Accuracy of the task. Given as a number from $0.0$ to $1.0$ with $1.0$ being the perfect score.
\end{itemize}

We also provide an anonymous summary of the questionnaire responses.

% \biblio
% \end{document}
%\documentclass[../supplemental.tex]{subfiles}

%\begin{document}

\section{Additional User Study Analysis}
\begin{figure*}[htb]
%\centering
  \includegraphics[width=1\textwidth]{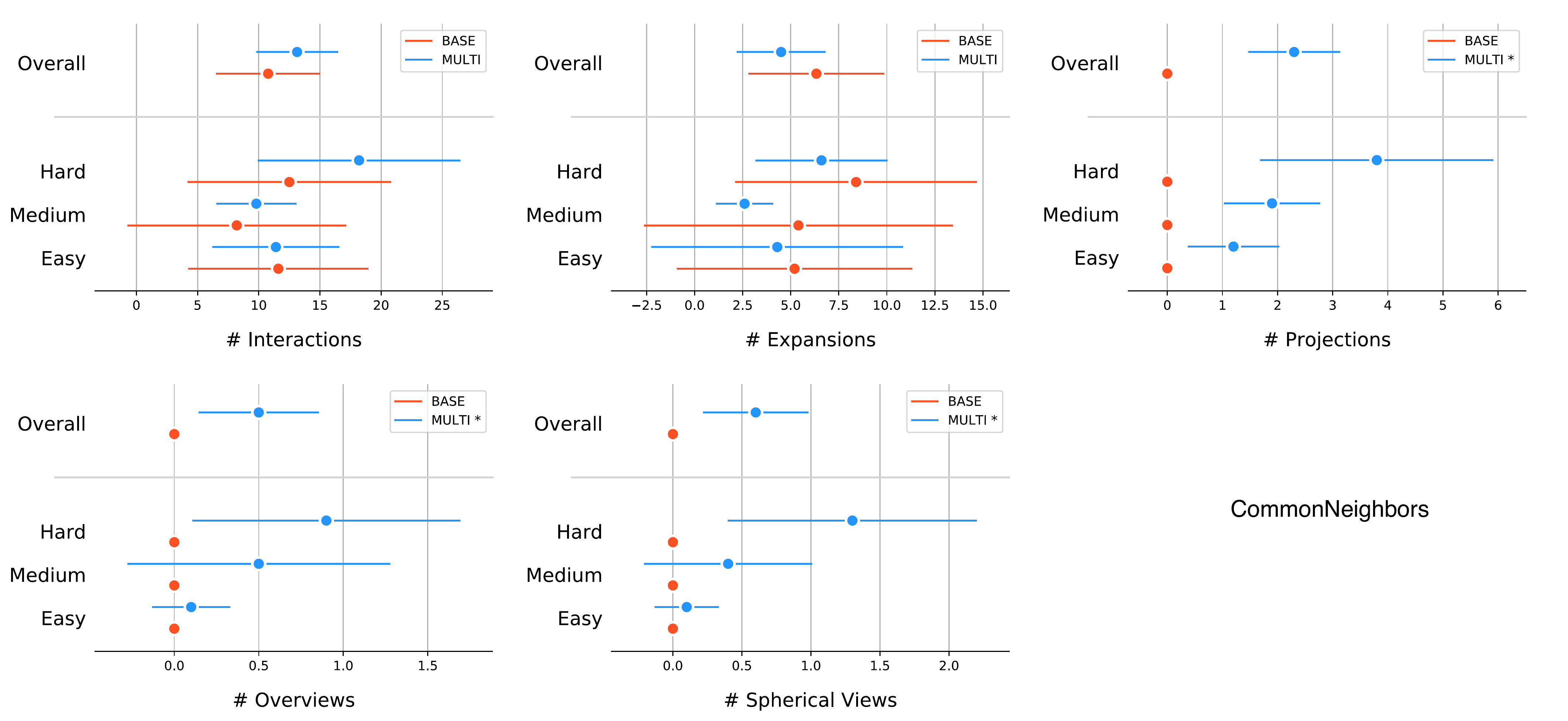}
  \caption{Performance results for the \textit{CommonNeighbors} task given as the mean over all participants. The error bars denote 95\% CIs. Each graph shows the cumulative number of interactions per type. '\# Interactions' is the sum of all other interaction types plus the number of task-specific selections made during the completion of the task. A star in the legend marks a significant difference in the overall results. Note that interaction counts for the base version are naturally zero for layouts that were not available in that condition.}
    % \vspace{-0.3in}
  \label{fig:ci_tasks_0}
\end{figure*}

\begin{figure*}[htb]
%\centering
  \includegraphics[width=1\textwidth]{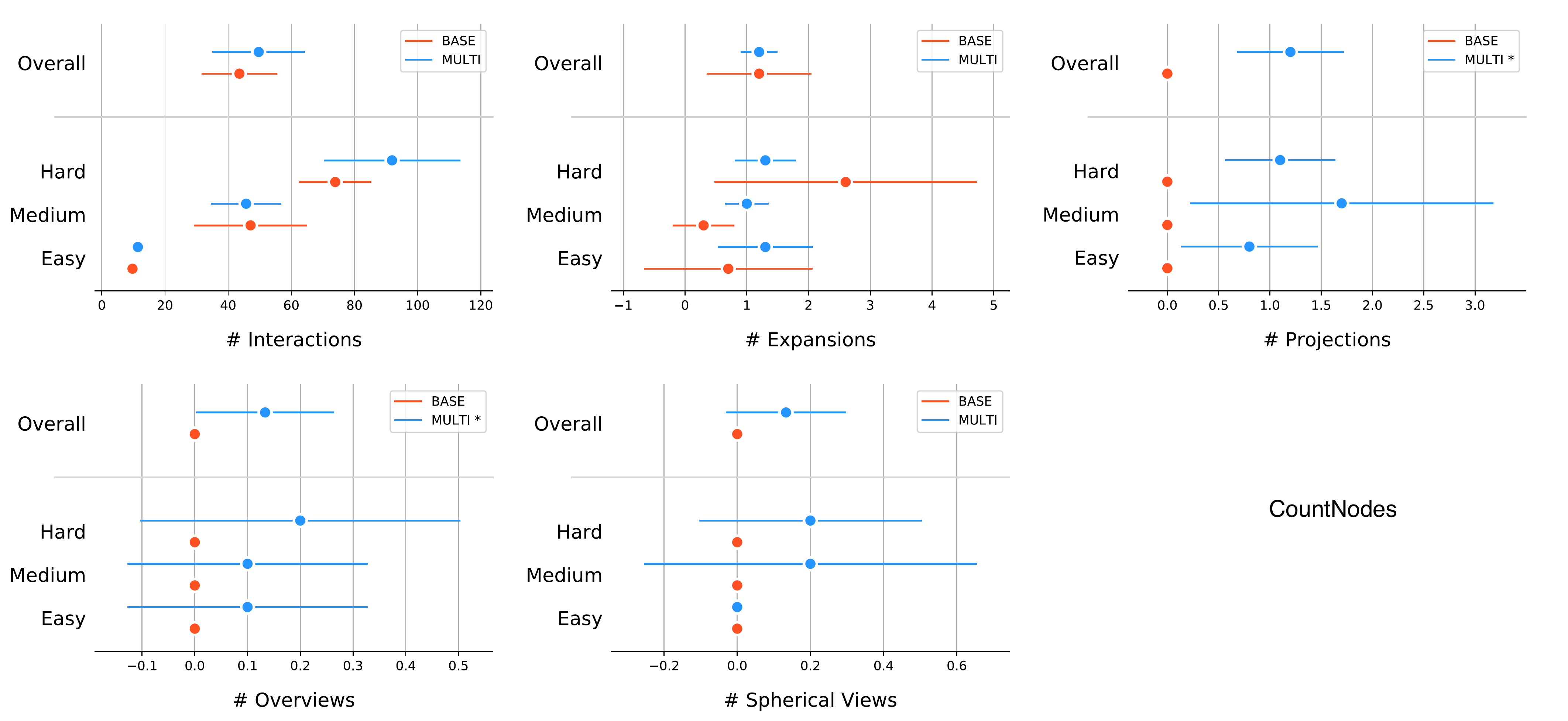}
  \caption{Performance results for the \textit{CountNodes} task given as the mean over all participants. The error bars denote 95\% CIs. Each graph shows the cumulative number of interactions per type. '\# Interactions' is the sum of all other interaction types plus the number of task-specific selections made during the completion of the task. A star in the legend marks a significant difference in the overall results.}
    % \vspace{-0.3in}
  \label{fig:ci_tasks_1}
\end{figure*}

\begin{figure*}[htb]
%\centering
  \includegraphics[width=1\textwidth]{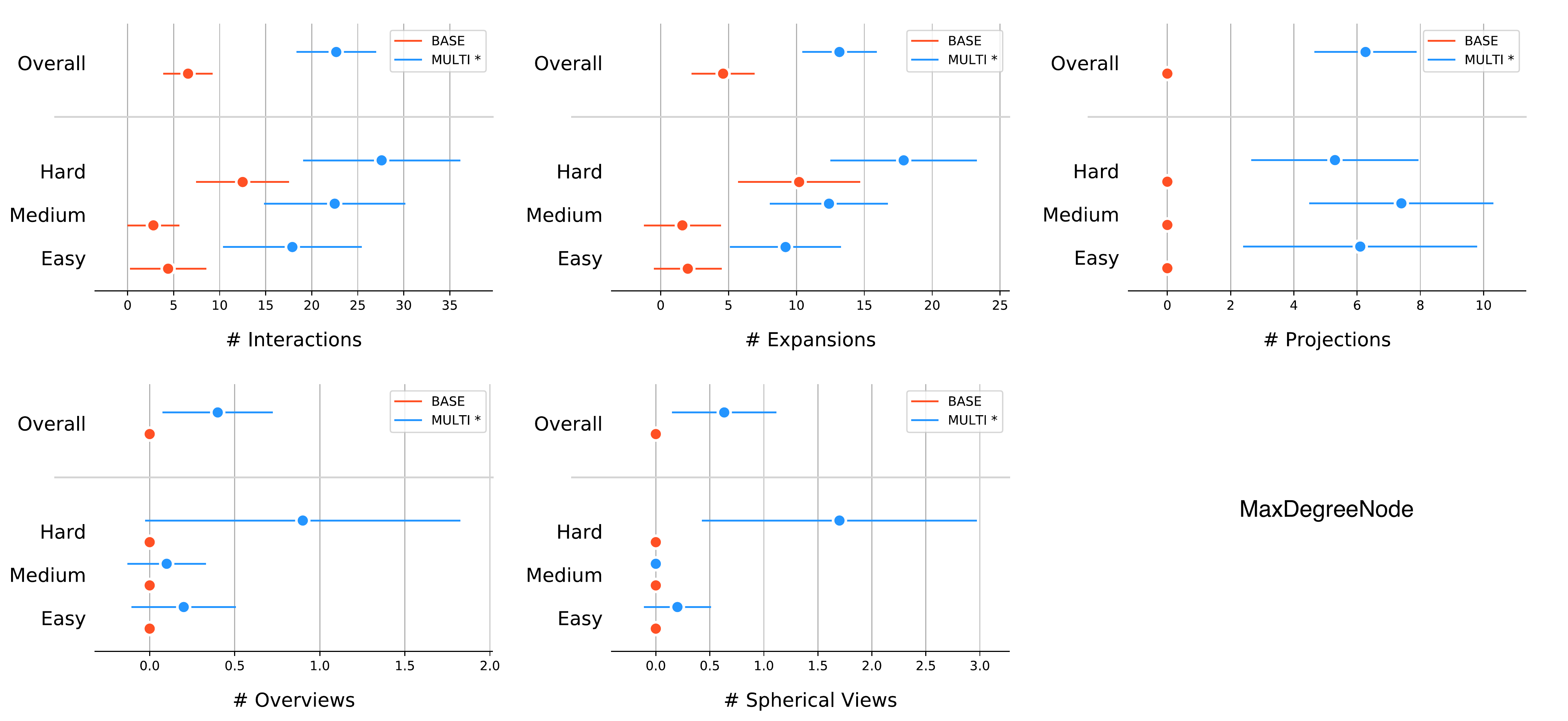}
  \caption{Performance results for the \textit{MaxDegreeNode} task given as the mean over all participants. The error bars denote 95\% CIs. Each graph shows the cumulative number of interactions per type. '\# Interactions' is the sum of all other interaction types plus the number of task-specific selections made during the completion of the task. A star in the legend marks a significant difference in the overall results.}
    % \vspace{-0.3in}
  \label{fig:ci_tasks_2}
\end{figure*}

\begin{figure*}[htb]
%\centering
  \includegraphics[width=1\textwidth]{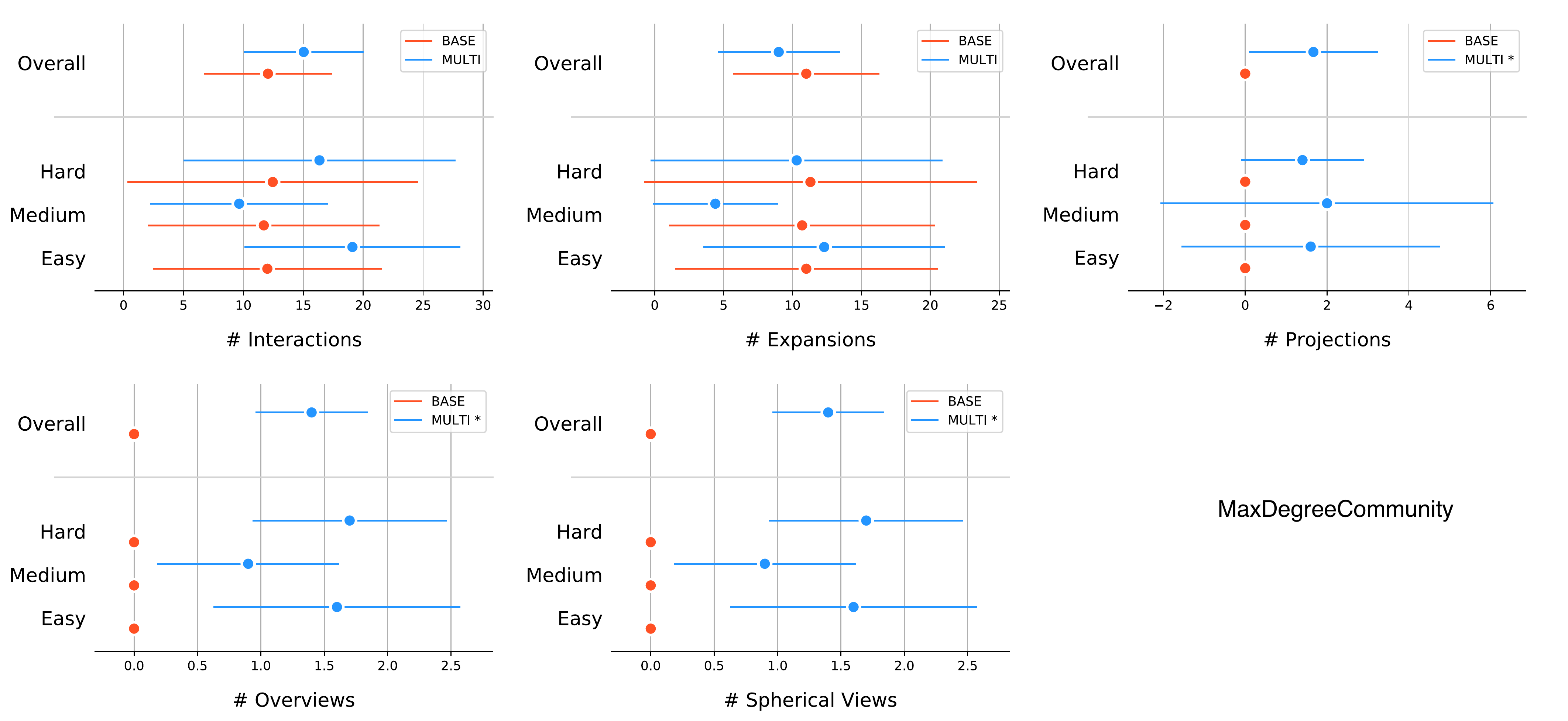}
  \caption{Performance results for the \textit{MaxDegreeCommunity} task given as the mean over all participants. The error bars denote 95\% CIs. Each graph shows the cumulative number of interactions per type. '\# Interactions' is the sum of all other interaction types plus the number of task-specific selections made during the completion of the task. A star in the legend marks a significant difference in the overall results.}
    % \vspace{-0.3in}
  \label{fig:ci_tasks_3}
\end{figure*}

\begin{figure*}[t]
%\centering
  \includegraphics[width=1\textwidth]{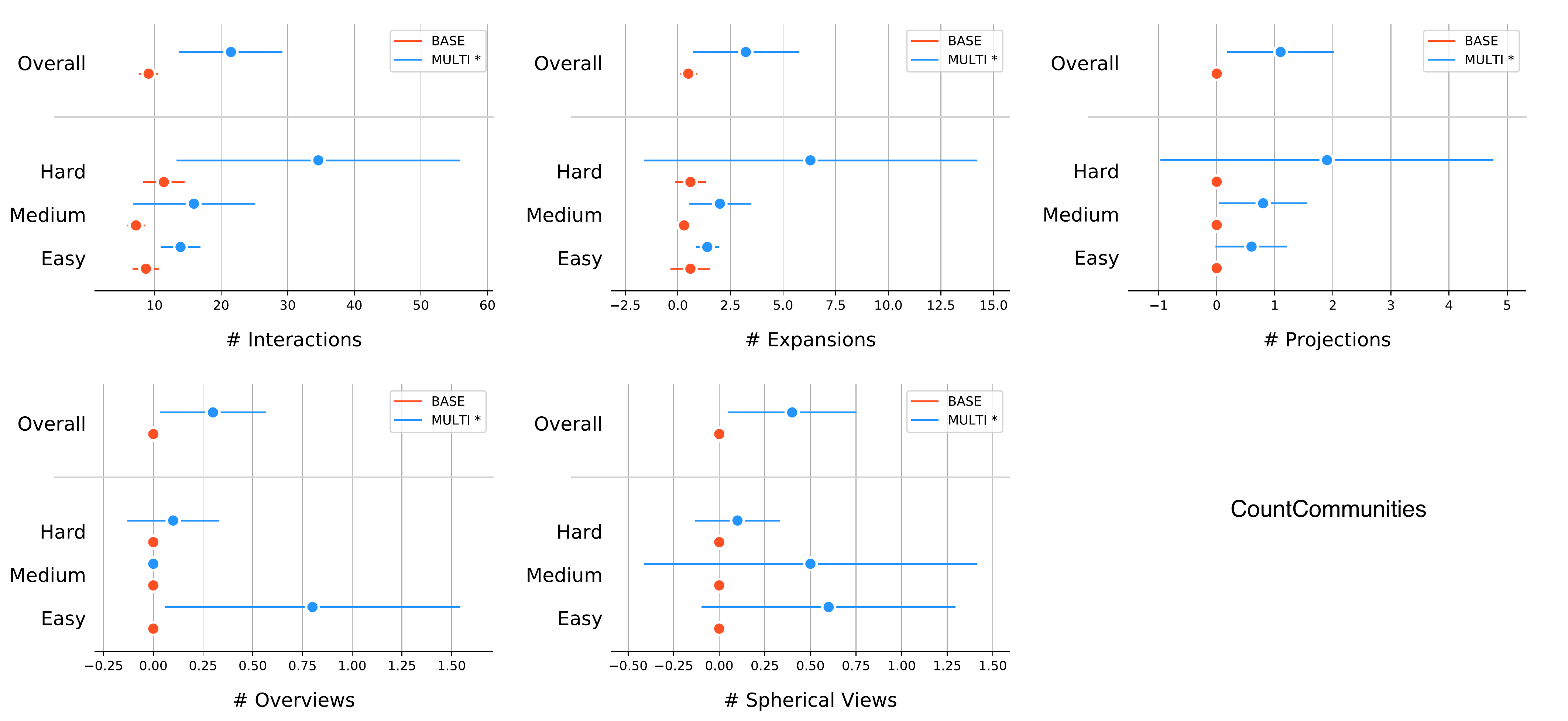}
  \caption{Performance results for the \textit{CountCommunities} task given as the mean over all participants. The error bars denote 95\% CIs. Each graph shows the cumulative number of interactions per type. '\# Interactions' is the sum of all other interaction types plus the number of task-specific selections made during the completion of the task. A star in the legend marks a significant difference in the overall results.}
    % \vspace{-0.3in}
  \label{fig:ci_tasks_4}
\end{figure*}

\begin{figure*}[t]
\centering
  \includegraphics[width=1.0\textwidth]{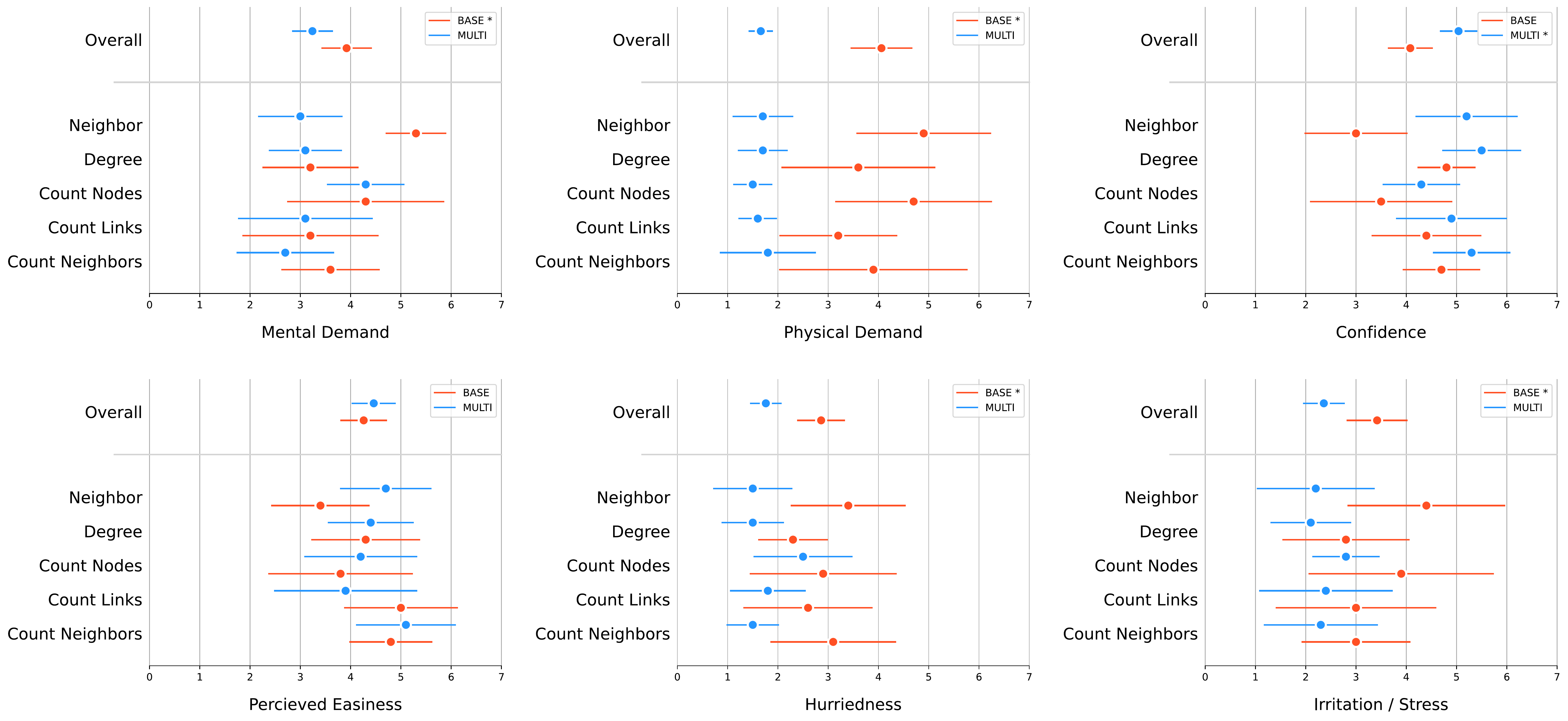}
  \caption{User's assessment of the tasks. Each graph shows the average results for one metric. The error bars denote 95\% CIs. Overall scores are shown at the top while scores per task category are displayed below. A star in the legend marks a significant difference in the overall results.}
    % \vspace{-0.3in}
  \label{fig:ci_feedback}
\end{figure*}

Here we provide additional evaluations of our user study. We list interaction counts for the different layouts split up per task and graph. Find them in Figures~\ref{fig:ci_tasks_0},~\ref{fig:ci_tasks_1},~\ref{fig:ci_tasks_2},~\ref{fig:ci_tasks_3}, and~\ref{fig:ci_tasks_4}. Additionally, we provide a more detailed analysis of user feedback using confidence intervals in Figure~\ref{fig:ci_feedback}.


% \biblio
% \end{document}

% \begin{acks}
% \end{acks}

% \bibliographystyle{ACM-Reference-Format}
